# Magnetization transfer from protons to quadrupolar nuclei in solid-state NMR using PRESTO or dipolar-mediated refocused INEPT methods

Raynald Giovine,[1] Julien Trébosc,[1] Frédérique Pourpoint,[1] Olivier Lafon,[1,2*] Jean-Paul Amoureux[1,3*]

[1] Univ. Lille, CNRS-8181, UCCS: Unit of Catalysis and Chemistry of Solids, F-59000 Lille, France.

[2] IUF, Institut Universitaire de France, 1 rue Descartes, 75231 Paris, France.

[3] Bruker France, 34 rue de l'Industrie, F-67166 Wissembourg, France.

**Abstract.** In solid-state NMR spectroscopy, the through-space transfer of magnetization from protons to quadrupolar nuclei is employed to probe proximities between those isotopes. Furthermore, such transfer, in conjunction with Dynamic Nuclear Polarization (DNP), can enhance the NMR sensitivity of quadrupolar nuclei, as it allows the transfer of DNP-enhanced $^1$H polarization to surrounding nuclei. We compare here the performances of two approaches to achieve such transfer: PRESTO (Phase-shifted Recoupling Effects a Smooth Transfer of Order), which is currently the method of choice to achieve the magnetization transfer from protons to quadrupolar nuclei and which has been shown to supersede Cross-Polarization under Magic-Angle Spinning (MAS) for quadrupolar nuclei and *D*-RINEPT (Dipolar-mediated Refocused Insensitive Nuclei Enhanced by Polarization Transfer) using symmetry-based $SR4_1^2$ recoupling, which has already been employed to transfer the magnetization in the reverse way from half-integer quadrupolar spin to protons.

We also test the PRESTO sequence with $R16_7^6$ recoupling using $270_0 90_{180}$ composite π-pulses as inversion elements. This recoupling scheme, which has previously been proposed to reintroduce $^1$H Chemical Shift Anisotropy (CSA) at high MAS frequencies with high robustness to rf-field inhomogeneity, has not so far been employed to reintroduce dipolar couplings with protons. These various techniques to transfer the magnetization from protons to quadrupolar nuclei are analyzed using (i) an average Hamiltonian theory, (ii) numerical simulations of spin dynamics, and (iii) experimental $^1$H → $^{27}$Al and $^1$H → $^{17}$O transfers in as-synthesized AlPO$_4$-14 and $^{17}$O-labelled fumed silica, respectively. The experiments and simulations are done at two magnetic fields (9.4 and 18.8 T) and several spinning speeds (15, 18-24 and 60 kHz). This analysis indicates that owing to its γ-encoded character, PRESTO yields the highest transfer efficiency at low magnetic fields and MAS frequencies, whereas owing to its higher robustness to rf-field inhomogeneity and chemical shifts, *D*-RINEPT is more sensitive at high fields and MAS frequencies, notably for protons exhibiting large offset or CSA, such as those involved in hydrogen bonds.

## I. Introduction

Quadrupolar nuclei, with nuclear spin quantum number $S > ½$, represent 75 % of stable NMR-active nuclei [1]. Numerous solids, such as organic compounds, biomolecules, hybrid or porous materials, nanoparticles, hydrates or heterogeneous catalysts, contain both quadrupolar nuclei and protons. For these materials, two-dimensional (2D) *D*-HETCOR (Dipolar-mediated HETeronuclear CORrelation) NMR experiments between quadrupolar nuclei and protons allow the unambiguous identification of proximities between sites occupied by these isotopes. Hence, these experiments facilitate the assignment of NMR spectra and provide precious information on the atomic-level structure of these materials. For instance, *D*-HETCOR experiments between $^{1}$H and $^{27}$Al isotopes have been employed to investigate the aluminum incorporation in aluminosilicate mesoporous material [2], the aluminum local environment in phyllosilicate [3], the dissolution mechanism of aluminosilicate glasses in water [4], the structure of aluminophosphates [5,6], aluminum-based metal-organic frameworks [7], alumina surfaces [8,9], olefin metathesis catalysts supported on chlorinated alumina support [10], methylaluminoxane-modified silica [11], the nature of Brønsted acid sites at the surface of amorphous silica alumina [12], and the location of Al atoms in zeolites [13,14]. Similarly, $^{1}$H-$^{11}$B *D*-HETCOR experiments have been employed to probe the changes in the local environment of boron atoms in borosilicate zeolites in the course of hydration and dehydration [15]. $^{1}$H-$^{43}$Ca *D*-HETCOR experiments have also been applied to observe the proximities between Ca atoms and hydroxyl groups in hydroxyapatite materials [16,17]. $^{1}$H-$^{17}$O *D*-HETCOR experiments have been employed to examine the structure of silica surfaces and silica-supported catalysts [18,19], the hydrogen bonds in crystalline and amorphous forms of pharmaceutical compounds [20], and crystalline peptides [21]. Furthermore, it has been demonstrated that the sensitivity gain provided by Dynamic Nuclear Polarization (DNP) enables the acquisition of $^{1}$H-$^{17}$O *D*-HETCOR 2D spectra for isotopically unmodified solids, despite the low natural abundance of $^{17}$O isotope [19,22,23]. Recently, $^{1}$H-$^{35}$Cl *D*-HETCOR experiments have been introduced to characterize the molecular-level structure of active pharmaceutical ingredients [24,25]. Besides half-integer spin quadrupolar nuclei, *D*-HETCOR experiments have been used for the indirect detection of $^{14}$N isotope which has a spin *S* = 1, via protons [26,27]. Such $^{1}$H-$^{14}$N experiments have been used to study the self-assembly of guanosine derivatives [28–30], the intermolecular hydrogen bonds and the nitrogen protonation in pharmaceuticals [30–35], the structure of layered aluminophosphate materials containing amine structure directing agents [5], and the host-guest interactions in metal-organic frameworks functionalized by amine groups [7]. *D*-HETCOR experiments can be achieved using either direct or indirect detection [36].

In direct detection, the magnetization of the excited nucleus is transferred to the detected one. Such transfer between spin-1/2 and quadrupolar isotopes under Magic-Angle Spinning (MAS) has first been performed using Cross-Polarization (CP) [37]. However, CP experiments that involve quadrupolar nuclei present numerous limitations when they are performed under MAS (CPMAS) [38,39]. First, the transfer efficiency is reduced since it is difficult to spin-lock the magnetization of quadrupolar nuclei for all crystallites simultaneously in a rotating powder [40]. Second, for half-integer spin quadrupolar nuclei, the most efficient CPMAS transfers are usually achieved for selective spin-lock of the central transition (CT) with a low radio-frequency (rf) field [39]. As a result, such transfers are then highly sensitive to resonance offset and Chemical Shift Anisotropy (CSA). Third, the optimization is difficult because the efficiency of the spin-lock for a half-integer spin quadrupolar isotope drops at the Rotary Resonance Recoupling (R$^3$) conditions; i.e. when the nutation frequency of the CT is a multiple of the MAS frequency, $\nu_R$ [41]. Fourth, CPMAS transfers are also sensitive to the strength of the quadrupole coupling constant, $C_Q$, and hence they may not be efficient for two sites exhibiting distinct $C_Q$ values [42].

Alternative *D*-HETCOR methods with direct detection have been introduced in order to circumvent the shortcomings of CPMAS transfers involving quadrupolar nuclei. These approaches include the *D*-RINEPT (Dipolar-mediated Refocused Insensitive Nuclei Enhanced by Polarization Transfer) [43–46] and PRESTO (Phase-shifted Recoupling Effects a Smooth Transfer of Order) [23,47–49] polarization transfers. The first introduced *D*-RINEPT experiment is TEDOR (Transferred-Echo DOuble Resonance) [43,50] using the REDOR (Rotational-Echo DOuble Resonance) scheme [51] as hetero-nuclear dipolar recoupling. However, REDOR does not eliminate the homo-nuclear dipolar couplings and hence is not suitable for *D*-HETCOR experiments with protons. More recently, *D*-RINEPT experiments, in which the hetero-nuclear dipolar couplings are reintroduced using the $R^3$ scheme, have been reported [44,45]. In particular, the $R^3$ scheme using an rf-field $\nu_1 = 2\nu_R$ has been employed to acquire *D*-RINEPT 2D spectra between protons and half-integer spin quadrupolar isotopes, such as $^{27}$Al or $^{17}$O [21,44]. Very recently, *D*-RINEPT experiments, in which heteronuclear couplings with protons are reintroduced using the symmetry-based $SR4_1^2$ recoupling scheme [52], have also been proposed to correlate protons with spin-1/2 nuclei with low gyromagnetic ratio, such as $^{83}$Y, $^{103}$Rh or $^{183}$W [53], or half-integer spin quadrupolar nuclei, such as $^{35}$Cl [46]. Nevertheless, to the best of our knowledge, such sequence has not yet been applied to transfer the magnetization from protons to half-integer spin quadrupolar nuclei.

In the PRESTO sequence, the hetero-nuclear dipolar couplings are reintroduced using symmetry-based single-quantum (1Q) hetero-nuclear γ-encoded dipolar recoupling schemes, such as $R18_1^7$ or $R18_2^5$ [23,47,48]. These symmetry-based sequences suppress the homo-nuclear dipolar interactions in the first-order average Hamiltonian. The sensitivity gain afforded by PRESTO has notably been used to transfer the DNP-enhanced proton polarization to $^{17}$O, without any sample labelling [23]. So far, the $RN_n^\nu$ schemes which have been used in PRESTO experiments employed single π-pulses as inversion element. Recently, schemes based on symmetries, such as $R20_9^8$, $R18_8^7$, $R14_6^5$, $R16_7^6$ and $R12_5^4$ and using $270_0 90_{180}$ composite π-pulses as inversion element, have been introduced to measure the $^1$H CSA at MAS frequencies of 60 and 70 kHz [54]. Here, the standard notation for composite pulses is used: $\xi_\varphi$ indicates a rectangular, resonant rf-pulse with flip angle ξ and phase φ, and the angles are written in degrees. These schemes reintroduce the same components of the spin interactions as the $RN_n^\nu$ schemes employed in PRESTO, but they benefit from higher robustness to rf-field inhomogeneity. However, to the best of our knowledge, these symmetry-based sequences have not yet been employed to reintroduce the heteronuclear dipolar interactions.

*D*-HETCOR experiments can also be used to increase the sensitivity for the NMR detection of half-integer spin quadrupolar isotopes. Such sensitivity gain has notably been reported when PRESTO scheme is used to transfer the DNP-enhanced proton polarization to quadrupolar nuclei, such as $^{17}$O and $^{27}$Al [23,48].

Indirect detection is an alternative to direct detection for *D*-HETCOR experiments; i.e. the excited isotope is also the detected one and the coherences are transferred back and forth between the isotopes. These indirectly detected *D*-HETCOR experiments particularly include the *D*-HMQC (Dipolar-mediated Hetero-nuclear Multiple-Quantum Correlation) schemes [44,55,56]. In these experiments, the hetero-nuclear dipolar couplings are reintroduced using various schemes, such as REDOR, $R^3$, SFAM (Simultaneous Frequency and Amplitude Modulation) and symmetry-based sequences [36,57–59]. In the case of *D*-HMQC experiments correlating protons and quadrupolar isotopes, the symmetry-based $SR4_1^2$ recoupling [52] is often employed. Indeed, this scheme: (i) exhibits high efficiency and robustness, (ii) is compatible with high MAS frequency, and (iii) can easily be optimized. We have also recently introduced another indirectly detected *D*-HETCOR experiment, called *D*-HUQC (Dipolar-mediated Hetero-nuclear Universal-Quantum Correlation), which employs γ-encoded symmetry-based recoupling schemes on the detected channel, and exhibits lower $t_1$-noise in the case of nuclei subject to large CSA [60].

The relative sensitivities of direct and indirect detections depend notably on the gyromagnetic ratios, the longitudinal relaxation times and the spectral widths of the correlated isotopes [46,61]. Furthermore, contrary to the directly detected *D*-HETCOR experiments, those using indirect detection cannot be used to acquire directly 1D spectra of quadrupolar isotope by transferring the DNP-enhanced polarization of protons to the nearby quadrupolar nuclei.

We focus here on the directly detected *D*-HETCOR experiments with proton excitation and detection of quadrupolar isotope. We compare the efficiency and the robustness of two techniques:

*D*-RINEPT using $SR4_1^2$ recoupling and PRESTO using simple $180_0$ pulses or $270_0 90_{180}$ composite ones. The two techniques are first described using an average Hamiltonian theory. Their performances are then compared using numerical simulations of spin dynamics and $^1$H-$^{27}$Al and $^1$H-$^{17}$O experiments on AlPO$_4$-14 and fumed silica, respectively.

## II. Pulse sequences and theory

### II.1. PRESTO

In the PRESTO sequence, the hetero-nuclear dipolar couplings between the protons and the quadrupolar nuclei are reintroduced under MAS by the application on the $^1$H channel of symmetry-based γ-encoded recoupling schemes, such as $R18_1^7$, $R18_2^5$, $R16_3^2$, $R18_4^1$ and $R16_7^6$. Those schemes recouple the $|m|$ = 2 space components and the one-quantum (1Q) terms of hetero-nuclear dipolar coupling ($^1$H-S) and $^1$H CSA (CSA$_H$) [47], while they suppress the $^1$H isotropic chemical shifts, the heteronuclear *J*-couplings with protons and the $^1$H-$^1$H dipolar couplings in the first-order average Hamiltonian. The rf-field requirements of $R18_1^7$, $R18_2^5$, $R16_3^2$, $R18_4^1$ and $R16_7^6$ recouplings with simple π-pulses are $\nu_1/\nu_R$ = 9, 4.5, 2.66, 2.25 and 1.14, respectively, whereas that of $R16_7^6$ scheme using $270_0 90_{180}$ composite π-pulses, denoted $R16_7^6$-C hereafter is 2.28. Schemes with high rf-field requirements, such as $R18_1^7$, may not be compatible with fast MAS. However, it must be noted that other γ-encoded recoupling schemes, with much lower rf-field requirements can be used [23].

During these recoupling schemes, the contribution of the dipolar coupling between *I* = $^1$H and *S* quadrupolar nuclei to the first-order average Hamiltonian is equal to [47]:

$$\overline{H}_{D,IS}^{(1)} = \omega_{D,IS} S_z [I^+ \exp(i2\varphi) + I^- \exp(-i2\varphi)] \quad (1)$$

where $I^\pm = I_x \pm iI_y$ are the shift operators. In Eq.1, the magnitude and phase of the recoupled *I-S* dipolar coupling are given by

$$\omega_{D,IS} = -\kappa \frac{\sqrt{3}}{2} b_{IS} \sin^2(\beta_{PR}^{D,IS}) \quad (2)$$

$$\varphi = \gamma_{PR}^{D,IS} + \alpha_{RL}^0 - \omega_R t^0 \quad (3)$$

In Eq.2, (i) $b_{IS}$ is the dipolar coupling constant in rad.s$^{-1}$, (ii) the dipolar scaling factor $\kappa$ = 0.182, 0.175, 0.161 and 0.152 for $R18_1^7$, $R18_2^5$, $R16_3^2$ and $R18_4^1$ schemes, respectively, with simple π-pulse as basic element and $\kappa$ = 0.15 for $R16_3^2$ scheme using $270_0 90_{180}$ composite π-pulse, which is denoted $R16_3^2$-C hereafter, and (iii) the Euler angles $\{0, \beta_{PR}^{D,IS}, \gamma_{PR}^{D,IS}\}$ relate the inter-nuclear *I-S* vector to the MAS rotor frame. In Eq.3, $\omega_R = 2\pi\nu_R$ and $t^0$ refers to the starting time of the symmetry-based scheme. The norm of $\overline{H}_{D,IS}^{(1)}$ does not depend on the $\gamma_{PR}^{D,IS}$ angle and hence, these recoupling schemes are called γ-encoded [45,62]. The recoupled Hamiltonian described in Eq.1 does not commute among different spin-pairs and the PRESTO experiment is hence affected by dipolar truncation, which may limit the observation of long *I-S* inter-nuclear distances. However, it must be noted that such dipolar truncation has been used to selectively correlate the signals of covalently bonded $^{13}$C and $^1$H nuclei [63,64].

In PRESTO, these heteronuclear dipolar recoupling schemes also reintroduce CSA$_H$, with the same scaling factor, and its first-order average Hamiltonian is equal to [47]:

$$\overline{H}_{CSA,I}^{(1)} = \omega_{CSA,I}^* I^+ + \omega_{CSA,I} I^- \quad (4)$$

where $\omega_{CSA,I}$ is the frequency of the recoupled CSA$_H$ and $\omega_{CSA,I}^*$ is its complex conjugate. This frequency is given by

$$\omega_{CSA,I} = -\frac{\kappa}{\sqrt{2}} [A_{22}^{CSA,I}]^R \exp\{-2i(\alpha_{RL}^0 - \omega_R t^0)\} \quad (5)$$

where $[A_{22}^{CSA,I}]^R$ is given by Eq.5 in ref.[36]. Eqs.1 and 4 show that the recoupled CSA$_H$ and S-$^1$H dipolar coupling terms do not commute. Therefore, the spin dynamics during PRESTO simultaneously depends on both CSA$_H$ and S-$^1$H hetero-nuclear dipolar coupling.

In the present article, we mainly employed the PRESTO-III variant, which is depicted in Fig.**1a** [47]. A π-pulse is applied at the centers of the two defocusing and refocusing periods, denoted τ and τ', respectively, and simultaneously the phase of the $^1$H channel irradiation is shifted by 180°. Such procedure limits the interference of $CSA_H$, because this interaction is only fully refocused when its tensor is axially symmetric and collinear with the S-H vector. During the τ delay, the longitudinal $^1$H magnetization is converted into $^1$H 1Q coherences, which are antiphase with respect to the *S* spin. The π/2 pulse on the *S* channel transforms the antiphase $^1$H 1Q coherences into *S* 1Q coherence antiphase with respect to $^1$H. A π-pulse is also applied at the center of the refocusing period, τ', in order to refocus the evolution under the isotropic shifts of the *S* nuclei, whereas the phase of the $^1$H channel irradiation is shifted by 180°. Furthermore, the τ/2 and τ'/2 delays must be integer multiples of the rotor period so that the evolution under $CSA_H$ and the second-order quadrupole interaction of the *S* nucleus is better refocused. Herein, we employed τ = τ'.

For a *S* spin coupled to a single proton with vanishing $CSA_H$, the PRESTO signal with τ = τ' is proportional to

$$S(\tau) \propto \langle \sin^2(\omega_{I,IS}\tau) \rangle = \frac{1}{2} - \frac{1}{3^{1/4}}\sqrt{\frac{\pi}{8\kappa b_{IS}\tau}}\left\{F_c\left(3^{1/4}\sqrt{\frac{2\kappa b_{IS}\tau}{\pi}}\right)\cos(\sqrt{3}\kappa b_{IS}\tau) + F_s\left(3^{1/4}\sqrt{\frac{2\kappa b_{IS}\tau}{\pi}}\right)\sin(\sqrt{3}\kappa b_{IS}\tau)\right\} \quad (6)$$

where the angular bracket ⟨…⟩ denotes the powder average. Eq.6 was derived using a closed analytical form for γ-encoded |m| = 1 recoupling sequences, and $F_c(x)$ and $F_s(x)$ are the Fresnel cosine and sine integrals, respectively [62]. This equation can be used for distance determination, in place of the spin dynamics simulations that have been employed in Ref [23]. In the absence of losses and $CSA_H$, the shorter τ value producing the maximal signal intensity is given by:

$$\text{PRESTO } (CSA_H = 0) \rightarrow \tau^{opt} = 2.18/(\kappa b_{IS}) \quad (7)$$

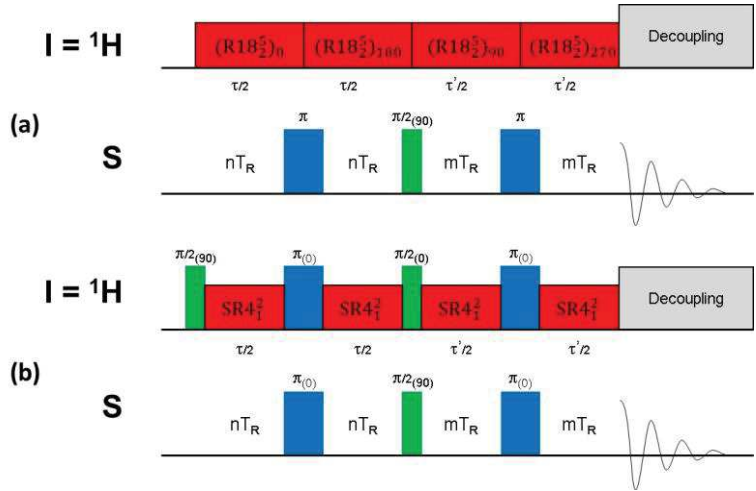

**Fig.1.** Pulse sequences for $I = {}^1H \rightarrow S$ transfers: (a) PRESTO-III-$R18_2^5$ and (b) *D*-RINEPT-$SR4_1^2$. The *S* isotope is quadrupolar with half-integer spin value and the *S* pulses are CT-selective. For the acquisition of *D*-HETCOR 2D spectra, the period $t_1$ is inserted (a) before the first $R18_2^5$ block and bracketed by two π/2-pulses, (b) between the first π/2-pulse and the first $SR4_1^2$ block. The quadrature detection along the indirect dimension was achieved using the States-TPPI procedure [65] by incrementing the phase of the first pulse prior to $t_1$ period. The phase cycling and pulse programs of the various sequences are given at the end of the SI.

## II.2. *D*-RINEPT

In *D*-RINEPT, described in Fig.**1b**, the S-$^1$H dipolar couplings are reintroduced using the $SR4_1^2$ recoupling [52]. This scheme is a 3-step multiple-quantum super-cycled version of $R4_1^2R4_1^{-2}$, each block lasting one rotor period: $T_R = 1/\nu_R$, i.e. $SR4_1^2 = [R4_1^2 R4_1^{-2}]_0 [R4_1^2 R4_1^{-2}]_{120} [R4_1^2 R4_1^{-2}]_{240}$, with $R4_1^2 = \pi_{90}\pi_{-90}\pi_{-90}\pi_{90}$ and $R4_1^{-2} = \pi_{-90}\pi_{90}\pi_{90}\pi_{-90}$, where $\pi_{90}$ and $\pi_{-90}$ denote resonant, rectangular π-pulses on $^1$H channel with phase y and -y, respectively. This sequence, during which the protons are irradiated with $\nu_1 = 2\nu_R$, reintroduces the CSA$_H$ and the $|m| = 1$ space component of the I-S dipolar coupling, whereas it suppresses the $^1$H isotropic chemical shifts, the $J_{IS}$-couplings, and the $^1$H-$^1$H dipolar couplings to the first-order. The $SR4_1^2$ scheme achieves zero-quantum hetero-nuclear dipolar recoupling and the contribution of the I-S dipolar coupling to the first-order average Hamiltonian is equal to [52,58]:

$$\bar{H}_{D,IS}^{(1)} = 2\omega_{D,IS} I_z S_z \tag{8}$$

$$\omega_{D,IS} = \frac{1}{4} b_{IS} \sin^2(\beta_{PR}^{D,IS}) \cos(2\varphi). \tag{9}$$

The norm of $\bar{H}_{D,IS}^{(1)}$ depends on the φ phase, given by Eq.3, and hence on the $\gamma_{PR}^{D,IS}$ angle. Therefore, the $SR4_1^2$ scheme is non-γ-encoded. The recoupled Hamiltonian of Eq.8 commutes among different spin-pairs, hence allowing the observation of long I-S inter-nuclear distances. Furthermore, the CSA$_H$ term recoupled by $SR4_1^2$ is proportional to $I_z$ and thus commutes with the I-S dipolar interaction of Eq.8. Hence, the CSA$_H$ does not interfere with the evolution under I-S dipolar interaction during *D*-RINEPT.

In the *D*-RINEPT sequence, the first π/2-pulse creates a transverse $^1$H magnetization in-phase with respect to the *S* spin. During the defocusing delay, τ, this magnetization evolves into transverse $^1$H magnetization antiphase with respect to the *S* spin. The simultaneous π-pulses on *S* and $^1$H channels at the center of the τ delay refocus the evolution under the CSA$_H$, while allowing that under the I-S dipolar interaction. Simultaneous π/2-pulses on the S and $^1$H channels convert the antiphase $^1$H magnetization into antiphase *S* one. During the refocusing delay τ′, this antiphase *S* magnetization is transformed into transverse in-phase *S* magnetization, which is detected during the acquisition period. The simultaneous π-pulses on *S* and $^1$H channels at the center of the τ′ delay refocus the evolution under the *S* isotropic chemical shifts, while allowing that under I-S dipolar couplings. For a proton coupled to a single *S* spin, the NMR signal of *D*-RINEPT-$SR4_1^2$ experiment with τ = τ′ is proportional to

$$S(\tau) \propto \frac{1}{2}\left\{1 - \frac{\pi\sqrt{2}}{4} J_{1/4}\left(\frac{b_{IS}}{4}\tau\right) J_{-1/4}\left(\frac{b_{IS}}{4}\tau\right)\right\} \tag{10}$$

where $J_{\pm 1/4}(x)$ denotes the Bessel functions of the first kind and ±1/4-order. In the absence of losses and CSA$_H$, the shorter τ value producing the maximal signal intensity is given by:

$$D\text{-RINEPT (CSA}_H = 0) \;\; \rightarrow \;\; \tau^{opt} = 9.44/b_{IS} \tag{11}$$

### III. Experimental section

#### III.1. Simulation parameters

All numerical simulations of spin dynamics were performed with the SIMPSON software (version 4.1.1) [66]. The powder average was calculated using 2304 {$\alpha_{MR}$, $\beta_{MR}$, $\gamma_{MR}$} Euler angles. The 256 {$\alpha_{MR}$, $\beta_{MR}$} angles, which relate the molecular and rotor frames, were selected according to the REPULSION algorithm [67], while the 9 $\gamma_{MR}$ angles were equally stepped from 0 to 360°. The dsyev method, with the corresponding Linear Algebra PACkage (LAPACK) [68], was used to accelerate the simulations [69]. During the PRESTO and *D*-RINEPT sequences, only CT-selective pulses are applied to the quadrupolar isotope. Therefore, its satellite transitions weakly contribute to the detected signal. This statement is supported by Fig.**S1**, showing that the simulated signal of *D*-RINEPT sequences for $^{13}$C-$^1$H$_4$ and $^{27}$Al-$^1$H$_4$ spin-systems are very similar, whereas the CPU time was 30-fold shorter for the former spin system than for the latter one. The ratio of the CPU times required for simulations on $^{13}$C-$^1$H$_4$ and $^{27}$Al-$^1$H$_4$ spin systems is consistent with the fact that the duration of the SIMPSON simulations is limited by the matrix-matrix multiplications and matrix diagonalizations. As the number of arithmetic operations for these processes scales with the cube of the dimension of the density matrix, a simulation for $^{13}C$-$^1H_4$ spin system should be ca. 27-fold faster than for $^{27}Al$-$^1H_4$ one [70]. Therefore, all simulations (except those of Fig.**S1b**) were performed for two spin-1/2 isotopes, $I$ = $^1H$ and $S$ = $^{13}C$, in order to accelerate the simulations.

The simulations were carried out for one isolated $^{13}C$-$^1H$ spin-pair, except those in Figs.**6-8**, **S1a**, **S6** and **S9**, which were carried for a $^{13}C$-$^1H_4$ spin-system in order to compare the robustness of the sequences to $^1H$-$^1H$ dipolar couplings, and those of Fig.**S1b**, which were carried for a $^{27}Al$-$^1H_4$ spin-system. In these five-spins systems, the four protons were located on the vertices of a tetrahedron and the $^{13}C$ or $^{27}Al$ nucleus was located on a symmetry axis of this tetrahedron. All $^1H$-$^1H$ dipolar coupling constants were identical and the $^{13}C$ or $^{27}Al$ nucleus was dipolar coupled with its closest proton with $|b_{IS}|/(2\pi)$ = 1 or 6 kHz. The CSA$_H$ value of the $^1H$ coupled to the $^{13}C$ or $^{27}Al$ nucleus is indicated in the figure captions, its asymmetry parameter is null, and the orientation of its principal axis systems with respect to the vector between its position and the $^{13}C$ or $^{27}Al$ nucleus is described by the Euler angles (0, 30°, 0).

The static magnetic field was fixed at $B_0$ = 18.8 T ($\nu_{0,1H}$ = 800 and $\nu_{0,13C}$ = 201 MHz) for all simulations, except for those of Fig.**7**, which were carried out at 9.4 T. The use of high magnetic fields is beneficial for half-integer spin quadrupolar nuclei, notably because the line-widths of the central transition are inversely proportional to $B_0$ and hence, the spectral resolution is proportional to $B_0^2$. In Figs.**2**-**6**, **8** and **S7**, the MAS frequency was 22 or 24 kHz (indicated as $\nu_R \approx$ 23 kHz), for R$18_2^5$ or SR$4_1^2$ schemes, respectively, to correctly sample the first maximum of the build-up curves with $|b_{IS}|/(2\pi)$ = 6 kHz. Such MAS frequencies correspond to those typically used for rotor with 3.2 mm outer diameter. In Figs.**S2** to **S9** (except Fig.**S7**), the MAS frequency was $\nu_R$ = 60 kHz for all recoupling sequences. This MAS frequency is accessible using rotor with an outer diameter of 1.3 mm and is generally required to achieve high resolution for the $^1H$ spectra without the use of $^1H$-$^1H$ dipolar decoupling sequence [71]. In Fig.**7**, the MAS frequency was 15 kHz to correspond to most DNP experiments.

We simulated the powder averaged signal of PRESTO-III-R$18_2^5$ and $D$-RINEPT-SR$4_1^2$ sequences, except in Fig.**7** where simulations were carried out for PRESTO-III-R$16_7^6$ using either single $\pi$-pulses or $270_0 90_{180}$ composite ones. The simulations carried out for PRESTO-II-R$18_2^5$ [not shown] confirm that this method is less robust to CSA$_H$ than the PRESTO-III variant. The PRESTO-III-R$18_2^5$ and $D$-RINEPT-SR$4_1^2$ sequences are denoted PRESTO and RINEPT hereafter. The pulses, which do not belong to the recoupling blocks were simulated as ideal Dirac pulses, except in Fig.**S1b**, where $^{27}Al$ CT-selective long pulses had to be used. The pulses of the recoupling schemes were applied on resonance, except in Fig.**3** and **S3**, for which the $^1H$ resonance offset was varied. For spin systems containing $^{13}C$, the transfer efficiencies of PRESTO and RINEPT were calculated as the ratios between the simulated signals of these experiments and that with a $^{13}C$ direct excitation with an ideal $\pi/2$-pulse. For Fig.**S1b**, the transfer efficiency of $^1H \to {}^{27}Al$ RINEPT experiment was calculated as the ratio of its simulated signal and that with a $^{27}Al$ direct excitation with a CT-selective $\pi/2$-pulse. Note that in SIMPSON simulations, the signal is not proportional to the gyromagnetic ratio, and hence the calculation of the transfer efficiency does not require to be normalized by the ratio of the gyromagnetic ratios. The build-up curves shown in Figs.**2** and **S2** were simulated using the shortest possible increments for $\tau/2$ = $\tau'/2$ delays, i.e. $2T_R/9$ for R$18_2^5$, corresponding to the length of a $\pi_{50}\pi_{-50}$ block, and $T_R/2$ for SR$4_1^2$, corresponding to the length of a $\pi_{90}\pi_{-90}$ block. The shortest possible increment for R$16_7^6$ and R$16_7^6$-C schemes is $7T_R/8$.

### III.2. Solid-state NMR experiments

For the experiments, all the rotors were fully packed. RINEPT experiments were acquired with SR$4_1^2$ recoupling scheme, whereas the PRESTO experiments were recorded with R$18_2^5$ and R$16_7^6$-C at $\nu_R$ = 15 kHz, R$18_2^5$ at $\nu_R$ = 20 kHz and R$16_3^2$, R$18_4^1$ and R$16_7^6$-C at $\nu_R$ = 60 kHz. $^1H \to {}^{27}Al$ PRESTO and RINEPT experiments were performed on an as-synthesized AlPO$_4$-14 sample with isopropylamine inserted into the pores [72], and the recycle delay was $\tau_{RD}$ = 1 s. The $^{27}$Al isotropic chemical shifts were referenced to 1 M solution Al(NO$_3$)$_3$, whereas the $^1$H isotropic chemical shifts were referenced to tetramethylsilane using the resonance of adamantane (1.74 ppm) as a secondary reference. $^1$H → $^{17}$O spectra were recorded on a fumed silica and the $^{17}$O isotropic chemical shifts were referenced to water at 0 ppm.

### III.2.a. Experiments at 9.4 T

Even if DNP-enhanced NMR experiments have been reported up to 21.1 T [73] and MAS frequency up to 40 kHz [74], DNP-enhanced PRESTO experiments have so far mainly been reported using 9.4 T and 3.2 mm rotors [23,48]. Therefore, we first recorded the $^1$H → $^{27}$Al PRESTO and RINEPT 2D experiments of AlPO$_4$-14 using a 9.4 T wide-bore magnet equipped with an Avance-II Bruker console. The experiments were recorded using a 3.2 mm HXY MAS probe used in the double resonance mode. The rf-fields of the pulses other than those used during the recoupling scheme were equal to 86 and 11 kHz on the $^1$H and $^{27}$Al channels, respectively.

### III.2.b. Experiments at 18.8 T

As mentioned above, high $B_0$ field is beneficial for the detection of quadrupolar nuclei. Therefore, PRESTO and RINEPT experiments transferring the magnetization of protons to quadrupolar nuclei were also performed on an 18.8 T narrow-bore Bruker magnet with HX double-resonance MAS probes. $^1$H → $^{27}$Al PRESTO and RINEPT experiments were performed on AlPO$_4$-14 using rotors with outer diameter of 3.2 and 1.3 mm, whereas $^1$H → $^{17}$O spectra were recorded with 3.2 mm diameter. With 3.2 mm rotors, experiments were recorded with an Avance III console, whereas we used an Avance IV one with 1.3 mm rotors.

For $^1$H → $^{27}$Al PRESTO and RINEPT experiments, the rf-field of the pulses other than those used during the recoupling scheme was $\nu_{1,^1H}$ = 77 and 208 kHz, and $\nu_{1,^{27}Al}$ = 10 and 14 kHz to achieve CT-selective excitation, at $\nu_R$ = 20 and 60 kHz, respectively.

$^1$H → $^{17}$O spectra were acquired at $\nu_R$ = 18 kHz on a fumed silica with specific surface area of 350 m$^2$/g, for which the surface was $^{17}$O enriched using a previously reported procedure [18]. The 1D $^{17}$O direct excitation MAS spectra were acquired using a single-pulse and QCPMG (quadrupolar Carr-Purcell-Meiboom-Gill) sequences [75]. Except during the recoupling parts on the proton channel, the rf-fields were $\nu_{1,^1H}$ = 100 and $\nu_{1,^{17}O}$ = 8 kHz.

### IV. Numerical simulations of the $^1$H → $^{13}$C transfer

### IV.1. Build-up curves at 18.8 T with single π-pulses

Fig.**2** shows the build-up curves of $^1$H → $^{13}$C PRESTO and RINEPT transfers for an isolated $^{13}$C-$^1$H spin pair at $\nu_R$ ≈ 23 kHz. For CSA$_H$ = 0, the PRESTO sequence exhibits stronger oscillations than RINEPT and a higher maximal transfer efficiency (0.73 for the former instead of 0.52 for the latter). Such differences are consistent with the γ-encoding recoupling used for PRESTO and the non-γ-encoding employed in RINEPT. For both sequences, the optimal recoupling times are in agreement with those predicted from Average Hamiltonian theory (Eqs.7 and 11). In addition, we can observe that the build-up curves of RINEPT are not affected by CSA$_H$, unlike those of PRESTO. This robustness of RINEPT to CSA$_H$ stems from its commutation with $^{13}$C-$^1$H dipolar terms recoupled by $SR4_1^2$, whereas those terms do not commute with $R18_2^5$ (see section II). Finally, the comparison of Figs.**2a** and **c** proves a larger influence of the CSA$_H$ on the PRESTO build-up curve in the case of small $^{13}$C-$^1$H dipolar couplings. It is noted that similar build-up curves are obtained for $\nu_R$ = 60 kHz (Fig.**S2**).

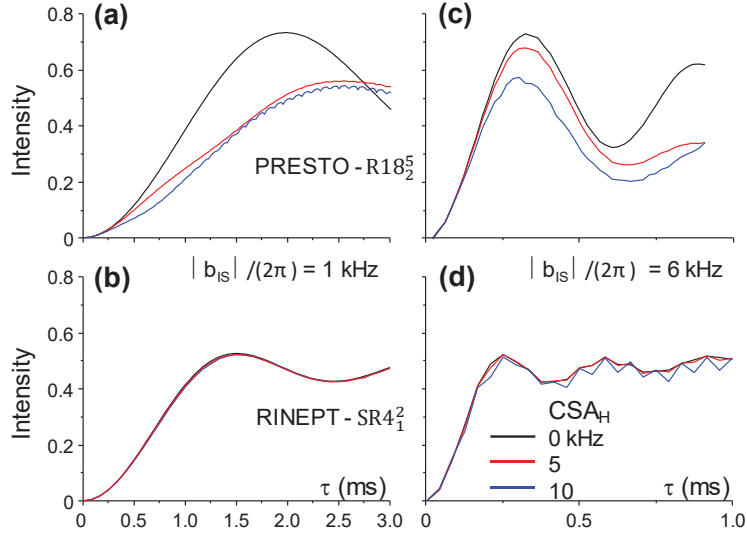

**Fig.2.** Simulated build-up curves of $^1H \rightarrow {}^{13}C$ transfer at 18.8 T and $\nu_R \approx 23$ kHz of (a,c) PRESTO-$R18_2^5$ with single $\pi$-pulses or (b,d) RINEPT for an isolated $^{13}C$-$^1H$ spin-pair with $|b_{IS}|/(2\pi)$ = (a,b) 1 or (c,d) 6 kHz, and CSA$_H$ = 0, 5 or 10 kHz.

### IV.2. Robustness to offset at 18.8 T with single $\pi$-pulses

The robustness of RINEPT is higher than that of PRESTO, especially for weak I-S dipolar couplings (Fig.**3**). Similar results are obtained at $\nu_R$ = 60 kHz (Fig.**S3**). This high robustness to offset of $SR4_1^2$ stems from the super-cycling, which better eliminates the unwanted cross-terms involving offset in the higher-order terms of the Average Hamiltonian. In the case of RINEPT, as expected by the Average Hamiltonian theory, the efficiency of the transfer versus the offset weakly depends on CSA$_H$. Furthermore, the robustness to $^1H$ offset improves for increasing MAS frequency since the rf-fields of the recoupling sequences are proportional to the MAS frequency (compare Figs.**3** and **S3**).

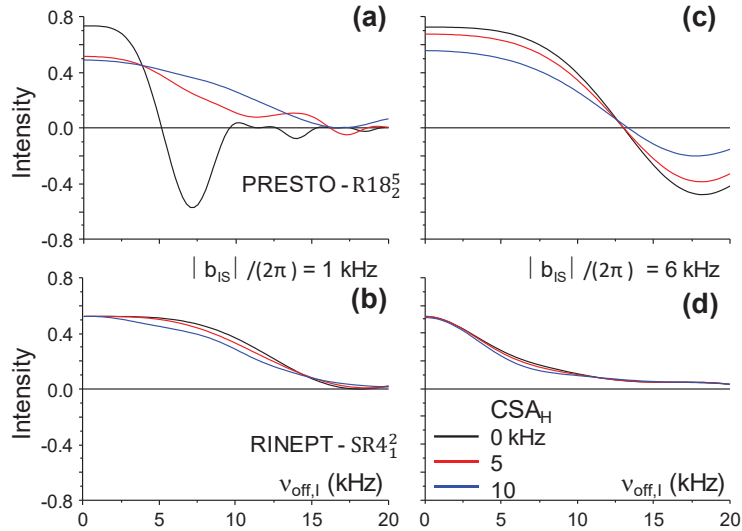

**Fig.3.** Simulated transfer efficiency versus the $^1H$ resonance offset, $\nu_{off,I}$, at 18.8 T and $\nu_R \approx 23$ kHz for (a,c) PRESTO-$R18_2^5$ with single $\pi$-pulses or (b,d) RINEPT, with $|b_{IS}|/(2\pi)$ = (a,b) 1 or (c,d) 6 kHz, and CSA$_H$ = 0, 5 or 10 kHz. The $\tau$ value was set to its optimum value determined from Fig.**2**.

### IV.3. Robustness to $^1H$ CSA at 18.8 T with single $\pi$-pulses

$^1$H CSA values can be as large as 30 ppm for protons involved in hydrogen bonds [76–78]. Such CSA corresponds to 24 kHz at 18.8 T. Figs.**4** and **S4** display a comparison of the robustness to CSA$_H$ of PRESTO and RINEPT transfers at $\nu_R \approx 23$ and 60 kHz, respectively. As already observed in Figs.**2** and **3** as well as **S2** and **S3**, the RINEPT method exhibits higher robustness to CSA$_H$ than PRESTO. This is consistent with the commutation between CSA$_H$ and hetero-nuclear dipolar coupling terms recoupled by $SR4_1^2$, whereas those terms do not commute with $R18_2^5$. As expected, the robustness to CSA$_H$ is improved at $\nu_R = 60$ kHz, as the rf-fields of the recoupling schemes are proportional to the MAS frequency.

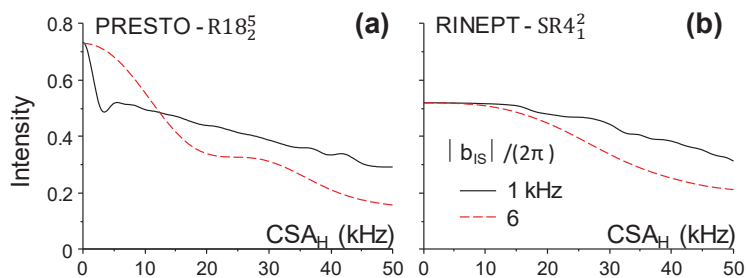

**Fig.4.** Simulated transfer efficiency versus CSA$_H$ at 18.8 T and $\nu_R \approx 23$ kHz for (a) PRESTO-$R18_2^5$ with single π-pulses and (b) RINEPT with $|b_{IS}|/(2\pi) = 1$ or 6 kHz. The τ value was set to its optimum value determined from Fig.**2**.

### IV.4. Robustness to rf-field inhomogeneity at 18.8 T with single π-pulses

The rf-field produced by a solenoid coil depends on the position inside the rotor [79–83]. It is known that for rotor diameters of 1.3 and 3.2 mm the minimal rf-field at the ends of the rotor is approximately 25 % of its maximum value [79,81]. In Figs.**5** and **S5**, the simulated transfer efficiencies of PRESTO and RINEPT experiments are plotted against the ratio between the applied and theoretical rf-fields, $R_{rf} = \nu_{1I}/\nu_{1,th}$. For both schemes, the transfer efficiency is below 10 % for $R_{rf} \leq 0.25$ [not shown]. Furthermore, these simulations show that the RINEPT scheme is much more robust to rf-inhomogeneity than the PRESTO one with single π-pulses. This result is attributed to the use of the $SR4_1^2$ recoupling, which is constructed from the amplitude-modulated basic sequences $R4_1^{\pm 2}$, i.e. the phase shift between consecutive π-pulses is 180°. This amplitude modulation achieves a compensation for rf-field errors [84,85].

These simulations indicate that the robustness to rf-inhomogeneity of $SR4_1^2$ does not depend on the $b_{IS}$ value, while that of PRESTO increases with this value. Actually, the difference $|\nu_{1I} - \nu_{1,th}|$ must be smaller than $|b_{IS}|/(2\pi)$ with $R18_2^5$, as already observed for the γ-encoded $R^3$ recoupling [45]. Therefore, at $\nu_R = 60$ kHz, the $R_{rf}$ relative interval yielding high transfer efficiency for PRESTO is smaller than at $\nu_R \approx 23$ kHz (compare Figs.**5a**,**b** and **S5a**,**b**) since $|\nu_{1I} - \nu_{1,th}| \leq |b_{IS}|$ and $\nu_{1,th}$ is proportional to $\nu_R$.

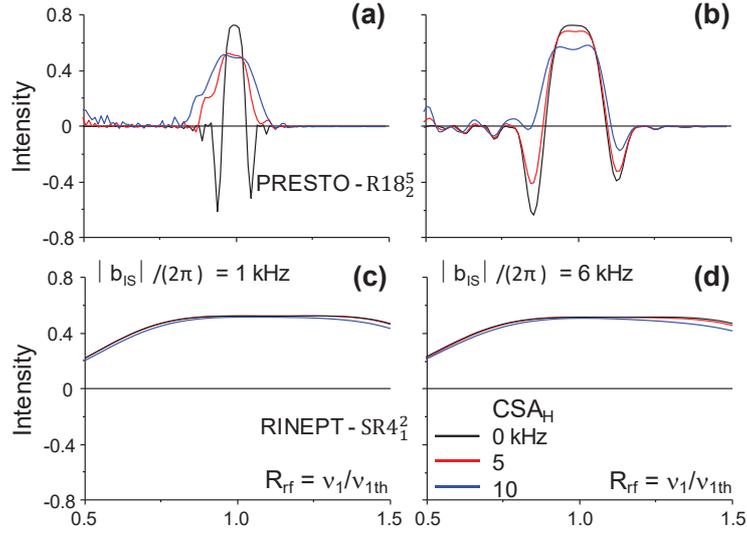

**Fig.5.** Simulated transfer efficiency at 18.8 T and $\nu_R \approx 23$ kHz versus the rf-inhomogeneity, $R_{rf} = \nu_1/\nu_{1th}$, for (a,b) PRESTO-R$18_2^5$ with single π-pulses or (c,d) RINEPT, with CSA$_I$ = 0, 5, 10 and $|b_{IS}|/(2\pi) = 1$ (a,c) or 6 (b,d) kHz. The τ value was set to its optimum value determined from Fig.**2**.

In Figs.**5** and **S5**, no $^1$H-$^1$H dipolar coupling interaction was considered. Figs.**6** and **S6** display the simulated robustness to rf-inhomogeneity of PRESTO and RINEPT sequences at $\nu_R \approx 23$ and 60 kHz, respectively, for the $^{13}$C-$^1$H$_4$ spin-system described in Section III.1, with $|b_{HH}|/(2\pi) = 0$, 1 or 7 kHz and $|b_{IS}|/(2\pi) = 1$ kHz. These simulations show that the robustness of PRESTO to rf-inhomogeneity does not depend on $b_{HH}$, whereas that of RINEPT decreases for increasing $^1$H-$^1$H dipolar interactions. The effect of these interactions on the RINEPT robustness does not depend on the MAS frequency (compare Figs.**6b** and **S6b**). Nevertheless, for the investigated spin systems with $|b_{HH}|/(2\pi)$ up to 7 kHz, RINEPT still exhibits higher robustness with respect to rf-inhomogeneity than PRESTO.

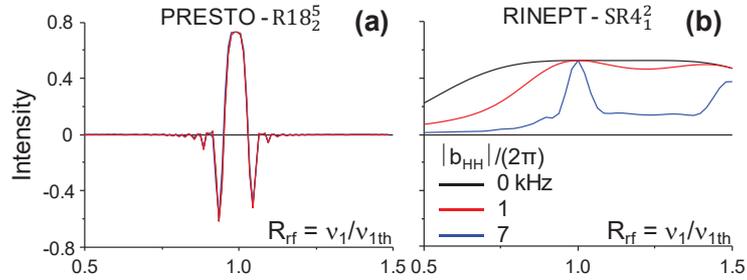

**Fig.6.** Simulated transfer efficiency at 18.8 T and $\nu_R \approx 23$ kHz versus the rf-inhomogeneity, $R_{rf} = \nu_1/\nu_{1th}$, for (a) PRESTO-R$18_2^5$ with single π-pulses or (b) RINEPT sequences applied to $^{13}$C-$^1$H$_4$ spin system with $|b_{HH}|/(2\pi) = 0$, 1 or 7 kHz, $|b_{IS}|/(2\pi) = 1$ kHz and CSA$_H$ = 0. The τ value was set to its optimum value determined from Fig.**2**.

### IV.5. Robustness to MAS frequency fluctuations at 18.8 T with single π-pulses

The transfer efficiencies of PRESTO and RINEPT versus the relative deviation, $R_{\nu_R} = (\nu_R - \nu_{R,th})/\nu_{R,th}$, of the actual MAS frequency from its theoretical value, $\nu_{R,th} \approx 23$ or 60 kHz, are plotted in Figs.**S7** or **S8**, respectively. As expected, the sensitivity to MAS fluctuations is higher for smaller hetero-nuclear dipolar coupling constants, which require longer recoupling times. Moreover, the absolute line-width of the efficiency curve only depends on $|b_{IS}|$ and therefore the relative deviation ($R_{\nu_R}$) is inversely proportional to the spinning speed (compare Figs.**S7** and **S8**). For $|b_{IS}|/(2\pi) = 6$ kHz, PRESTO and RINEPT experiments exhibit similar robustness to MAS fluctuations. However, for $|b_{IS}|/(2\pi) = 1$ kHz, this robustness decreases for increasing CSA$_H$ in the case of RINEPT, whereas it increases in the case of PRESTO. Hence, in the case of small dipolar couplings between *S*-spin and protons subject to significant CSA$_H$, PRESTO is more robust to MAS fluctuations than RINEPT. However, even in that case, the simulations show that the stability of the MAS frequency achieved using the latest generation of MAS speed controllers is sufficient to avoid significant intensity losses for both PRESTO and RINEPT.

### IV.6. Robustness to $^1$H-$^1$H dipolar interactions

Protons in solids, notably in organic and hybrid ones, are often coupled to several other protons. Therefore, the $^1$H→$^{13}$C transfer efficiency was simulated for the $^{13}$C-$^1$H$_4$ spin system (described in section III.1) versus the $^1$H-$^1$H dipolar coupling constant, $|b_{HH}|/(2\pi)$. The results are shown in Figs.**7**, **8** and **S9**, for $\nu_R$ = 15, 23 and 60 kHz, respectively. For the three MAS frequencies and for both PRESTO and RINEPT schemes, the effect of $^1$H-$^1$H dipolar couplings is larger for smaller $|b_{IS}|/(2\pi)$ values due to longer recoupling times. For the three MAS frequencies, PRESTO-$R18_2^5$ and RINEPT exhibit similar robustness to $^1$H-$^1$H dipolar couplings, whereas $SR4_1^2$ recoupling employs an rf-field 2.25-fold smaller than that of $R18_2^5$. As seen in Fig.**7**, $R16_7^6$ recoupling, which employs an rf-field 40% smaller than that of $SR4_1^2$ and 4-fold smaller than that of $R18_2^5$, is much more sensitive to $^1$H-$^1$H dipolar couplings than these schemes. The replacement of single π-pulses by $270_090_{180}$ ones increases both the required rf-field and the robustness to $^1$H-$^1$H dipolar couplings. However, $R16_7^6$-C recoupling is more sensitive to $^1$H-$^1$H dipolar couplings than $R18_2^5$ and $SR4_1^2$ schemes, even if the rf-field of $R16_7^6$-C is 28% higher than that of $SR4_1^2$. The comparison of Figs.**7**, **8** and **S9** also shows that the use of high MAS frequency improves the robustness to $^1$H-$^1$H dipolar interactions for both PRESTO and RINEPT experiments.

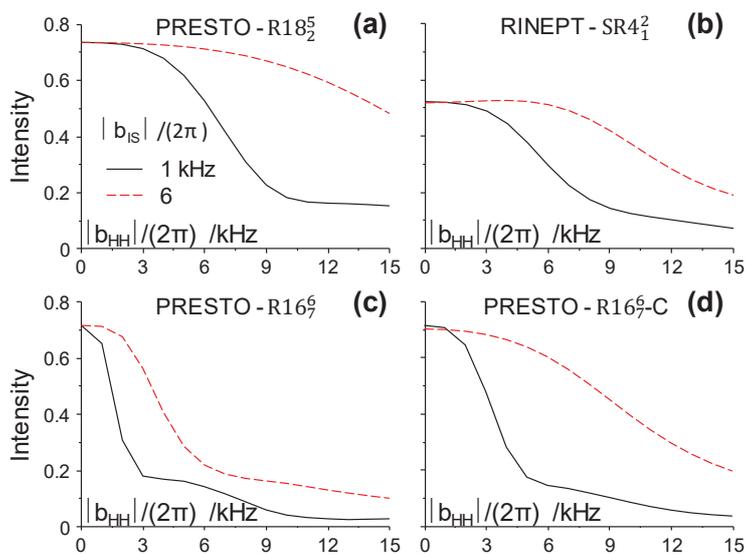

**Fig.7.** Simulated on-resonance transfer efficiency at 9.4 T with $\nu_R$ = 15 kHz, CSA$_H$ = 0 and $|b_{IS}|/(2\pi)$ = 1 or 6 kHz, versus $|b_{HH}|/(2\pi)$ in $^{13}$C-$^1$H$_4$ spin system for (a,c,d) PRESTO with $R18_2^5$ (a) and $R16_7^6$ (c) with single π-pulses, or $R16_7^6$-C with composite π-pulses (d), as well as (b) RINEPT. The τ value were set to 2, 1.46, 4.67 and 2.22 ms for subfigures a, b, c and d, respectively.

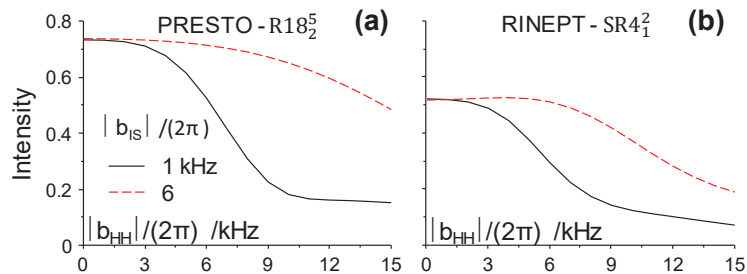

**Fig.8.** Simulated on-resonance transfer efficiency at 18.8 T and $\nu_R \approx 23$ kHz versus the $|b_{HH}|/(2\pi)$ constant in $^{13}C$-$^1H_4$ spin system for (a) PRESTO-R$18_2^5$ with single π-pulses and (b) RINEPT schemes with CSA$_H$ = 0 and $|b_{IS}|/(2\pi)$ = 1 or 6 kHz. The τ value was set to its optimum value determined from Fig.**2**.

### V. NMR experiments

#### V.1. Experiments on AlPO$_4$-14 at 9.4 T with $\nu_R$ = 15 kHz

The crystal structure of AlPO$_4$-14 exhibits four crystallographically inequivalent Al sites: two AlO$_4$, one AlO$_5$ and one AlO$_6$, with C$_Q$ = 1.8, 4.1, 5.6 and 2.6 MHz, respectively [86]. Even at 18.8 T, the two AlO$_4$ resonances overlap, and thus only their sum signal is given in Tables **1** to **3**. After optimization of the τ delay and the rf-field of the recoupling scheme [not shown], we recorded three *D*-HETCOR 2D spectra with $^{27}$Al detection, denoted $^{27}$Al-{$^1$H} hereafter, of AlPO$_4$-14 using $^1$H → $^{27}$Al PRESTO-R$18_2^5$ (Fig.**9a**), PRESTO-R$16_7^6$-C and RINEPT sequences. The $^1$H dimension of the 2D spectra exhibits three resolved proton signals, NH$_3^+$, CH and CH$_3$. For the three 2D spectra, all $^1$H-$^{27}$Al cross-peaks were detected. As seen in Table 1, the signal-to-noise ratio (S/N) of the cross-peaks of PRESTO-R$18_2^5$ is in average 50% higher than that of RINEPT. The higher sensitivity of PRESTO-R$18_2^5$ stems from the γ-encoding of R$18_2^5$ schemes, which results in higher transfer efficiency, whereas the SR$4_1^2$ scheme is non-γ-encoded (Fig.**2**). Nevertheless, the S/N ratio of the RINEPT 2D spectrum is in average about 73% higher than that of PRESTO-R$16_7^6$-C. The lower sensitivity of PRESTO-R$16_7^6$-C must stem from its lower robustness to $^1$H-$^1$H dipolar couplings (Fig.**7**).

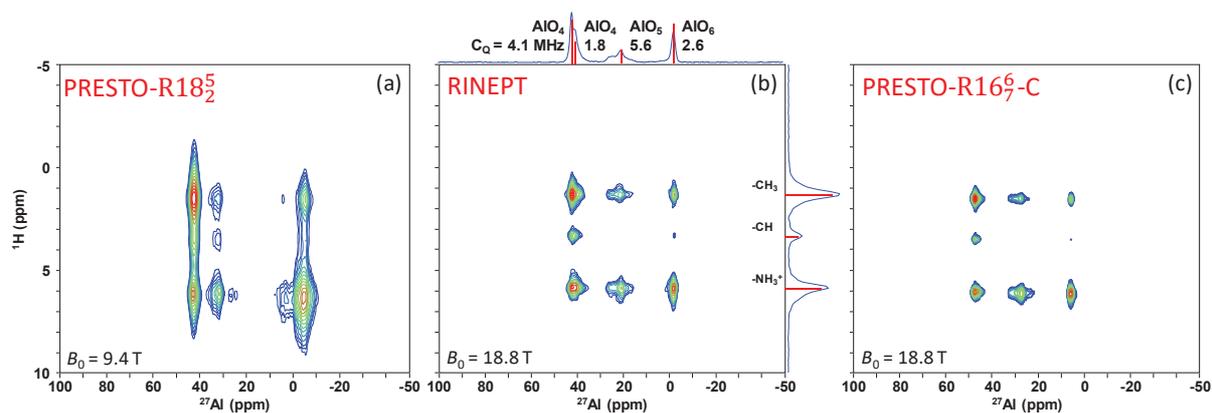

**Fig.9.** $^{27}$Al-{$^1$H} 2D spectra of AlPO$_4$-14 acquired with τ = 800 μs, using the following conditions: (a) $\nu_R$ = 15 kHz, 9.4 T with PRESTO-R$18_2^5$ and single π-pulses, (b) $\nu_R$ = 20 kHz, 18.8 T with RINEPT, (c) $\nu_R$ = 60 kHz, 18.8 T with PRESTO-C-R$16_7^6$. The $^1$H rf-field during the recoupling, the number of scans, the number of $t_1$ increments, and the total experimental time, were equal to ($\nu_1$ (kHz), NS, N$_1$, T$_{exp}$) = (68, 128, 100, 3.5 h) for (a), (46, 4, 128, 10 min) for (b) and (140, 128, 128, 4.5 h) for (c). The assignment of $^1$H and $^{27}$Al signals is shown on the projections: The C$_Q$ values are indicated on top of the $^{27}$Al projection of the spectrum (b).

**Table 1.** S/N ratios at 9.4 T with AVANCE-II console and $\nu_R$ = 15 kHz of the cross-peaks in $^{27}$Al-{$^1$H} 2D HETCOR spectra of AlPO$_4$-14 acquired with RINEPT, PRESTO-R$18_2^5$ with single π-pulse, and PRESTO-R$16_7^6$-C.

| Sequence | PRESTO-R$18_2^5$ | | | RINEPT | | | PRESTO-R$16_7^6$-C | | |
|---|---|---|---|---|---|---|---|---|---|
| $\delta_{iso,^{27}Al}$ | 42 | 22 | -2 | 42 | 22 | -2 | 42 | 22 | -2 |
| $\delta_{iso,^1H}$ | Al$^{IV}$ | Al$^V$ | Al$^{VI}$ | Al$^{IV}$ | Al$^V$ | Al$^{VI}$ | Al$^{IV}$ | Al$^V$ | Al$^{VI}$ |
| 1.3 (CH$_3$) | 68 | 37 | 65 | 44 | 25 | 34 | 27 | 16 | 17 |
| 3.3 (CH) | 43 | 15 | 20 | 28 | 10 | 10 | 16 | 6 | 5 |
| 5.8 (NH$_3^+$) | 95 | 17 | 32 | 81 | 16 | 25 | 52 | 9 | 15 |

**V.2. Experiments on AlPO$_4$-14 at 18.8 T with single π-pulses and $\nu_R$ = 20 kHz.**

In order to test the influence of the $B_0$ field on the sensitivity, $^1$H → $^{27}$Al PRESTO-R$18_2^5$ and RINEPT experiments were carried out on AlPO$_4$-14 at $B_0$ = 18.8 T with $\nu_R$ = 20 kHz. Figs.**10b**,**e** show the build-up curves of the four $^{27}$Al signals obtained with $^1$H → $^{27}$Al PRESTO and RINEPT. The experimental build-up curves significantly differ from the simulated ones of Fig.**2**. In particular, it is noted that the build-up curve of PRESTO exhibits smaller oscillations than the simulated curves. Furthermore, the experimental optimal recoupling time is τ ≈ 800 μs for PRESTO and RINEPT, whereas according to average Hamiltonian theory and numerical simulations for an isolated spin-pair, the optimal recoupling time of PRESTO-R$18_2^5$ should be 32 % longer than that of RINEPT (Eqs.7 and 11). The discrepancy between simulations and experiments may be attributed to the presence of several protons in the sample instead of isolated spin-pairs (Fig.**9**), as well as the coherent and incoherent losses during the τ delays.

The $^1$H → $^{27}$Al PRESTO and RINEPT signal intensity as function of the rf-field of the recoupling scheme is shown in Figs.**10a**,**d**. The maximal signal intensity is achieved for nutation frequencies close to the theoretical ones: $\nu_1 \approx 4.5\nu_R$ for PRESTO and $2\nu_R$ for RINEPT. Furthermore, in agreement with simulations (Fig.**5**), the RINEPT-SR$4_1^2$ recoupling is more robust to rf-inhomogeneity than PRESTO with single π-pulses. The intervals of rf-field, for which the signal intensity is larger than half of its maximal value, are equal to 15 and 40 kHz for PRESTO and RINEPT, respectively, which correspond to 17 and 100 %, in relative value.

Furthermore, as seen in Figs.**10c**,**f** the PRESTO sequence, which employs γ-encoded R$18_2^5$ recoupling, is slightly less sensitive to the MAS frequency than RINEPT, which uses the non-γ-encoded SR$4_1^2$ scheme. However, modern speed controllers can achieve stability much better than ±20 Hz. In such interval of MAS frequency, both PRESTO and RINEPT sequences are insensitive to MAS frequency fluctuations.

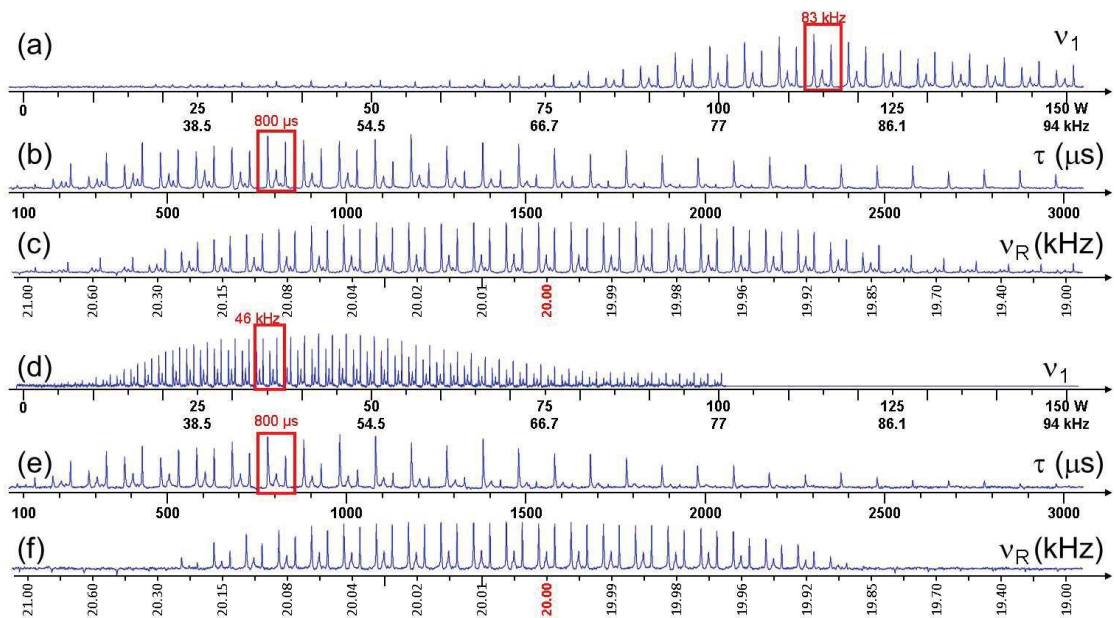

**Fig.10.** $^1$H → $^{27}$Al (a-c) PRESTO-R$18_2^5$ with single π-pulses and (d-f) RINEPT spectra of AlPO$_4$-14 at $B_0$ = 18.8 T versus $\nu_1$ (a,d), τ (b,e) and $\nu_R$ (c,f) using $\nu_R$ = 20 kHz in (a,b,d,e), τ = 800 μs in (a,c,d,f), $\nu_1$ = 83 (b,c) or 46 kHz (e,f). Each spectrum was recorded using *NS* = 32.

Fig.**9b** shows the $^{27}$Al-{$^1$H} RINEPT 2D spectrum of AlPO$_4$-14 at $B_0$ = 18.8 T with $\nu_R$ = 20 kHz. As expected, it exhibits a higher resolution along the $^{27}$Al dimension than in Fig.**9a**, since for half-integer quadrupolar nuclei, the resolution is proportional to $B_0^2$. Furthermore, the $^{27}$Al isotropic shifts are distinct between 9.4 and 18.8 T since the quadrupolar induced shifts are inverse-proportional to $B_0$. For both $^{27}$Al-{$^1$H} RINEPT and PRESTO-R$18_2^5$ 2D spectra, all $^{27}$Al-$^1$H cross-peaks were detected, even if the (AlO$_5$, CH) cross-peak exhibits a small intensity and is not visible in Fig.**9b**. As seen in Table 2, the cross-peaks of the PRESTO-R$18_2^5$ spectrum exhibit a S/N ratio in average 36% higher than for the RINEPT spectrum acquired within an identical experimental time. The sensitivity gain for PRESTO-R$18_2^5$ method with respect to RINEPT scheme decreases with increasing $B_0$ field. Such decrease stems notably from the larger $^1$H offset and CSA at high field.

**Table 2.** S/N ratios at 18.8 T with AVANCE-III console and $\nu_R$ = 20 kHz of the cross-peaks in $^{27}$Al-{$^1$H} 2D HETCOR spectra of AlPO$_4$-14 acquired with RINEPT and PRESTO-R$18_2^5$ with single π-pulses.

| Sequence | PRESTO-R$18_2^5$ | | | RINEPT | | |
|---|---|---|---|---|---|---|
| $\delta_{iso,^{27}Al}$ | 42 | 22 | -2 | 42 | 22 | -2 |
| $\delta_{iso,^1H}$ | Al$^{IV}$ | Al$^V$ | Al$^{VI}$ | Al$^{IV}$ | Al$^V$ | Al$^{VI}$ |
| 1.3 (CH$_3$) | 83 | 14 | 35 | 61 | 11 | 20 |
| 3.3 (CH) | 23 | 4 | 7 | 18 | 3 | 4 |
| 5.8 (NH$_3^+$) | 66 | 24 | 72 | 65 | 21 | 53 |

### V.3. Experiments on AlPO$_4$-14 at 18.8 T with $\nu_R$ = 60 kHz.

It can be desirable to transfer the magnetization of protons to half-integer spin quadrupolar nuclei at fast MAS, which improves the $^1$H resolution by averaging out the $^1$H-$^1$H dipolar couplings [71]. Fast MAS also enhances by a factor of 3 to 4 the spectral resolution of half-integer spin quadrupolar nuclei subject to large quadrupole interactions by separating the spinning sidebands from the center-band [87]. As the rf-field requirement of R$18_2^5$, $\nu_1 \approx 4.5\nu_R$ with single π-pulses, is incompatible with the rf-specifications of most 1.3 mm MAS probes, PRESTO experiments were carried out using $R16_3^2$, $R18_4^1$ and $R16_7^6$-C recoupling schemes, which only require $\nu_1/\nu_R \approx 2.66$, 2.25 and 2.28, respectively.

We acquired the build-up curves of RINEPT and PRESTO experiments at $\nu_R$ = 60 kHz [not shown] and found optimal recoupling times of $\tau \approx 800$ µs. This is the same value as that obtained for the experiments performed at $\nu_R$ = 20 kHz. Fig.**11** shows the $^1$H → $^{27}$Al PRESTO and RINEPT signals of AlPO$_4$-14 versus $\nu_1$. The full widths at half maximum of NMR signal intensity as function of the rf-field are similar (ca. 20 kHz) for $R16_3^2$ and $R18_4^1$ schemes at $\nu_R$ = 60 kHz and for $R18_2^5$ at $\nu_R$ = 20 kHz, whereas they are two-fold broader at $\nu_R$ = 60 than at 20 kHz for RINEPT (compare Figs. **10a**,**d** and **11**). Hence, the tolerated relative deviation of the rf-field is about ±6% for $R18_4^1$ and $R16_3^2$ and ±20% for RINEPT at $\nu_R$ = 60 kHz, instead of ±12% for $R18_2^5$ and ±30% for RINEPT at $\nu_R$ = 20 kHz. These experimental results are consistent with the simulations shown in Figs.**5** and **S5** and they indicate that the PRESTO sequence with single π-pulses is more sensitive to the rf-field homogeneity at high MAS frequency than RINEPT since the rf-field inhomogeneity corresponds to a relative variation of the rf-field amplitude in the sample space. On the contrary, PRESTO with composite π-pulses is much more robust to rf-inhomogeneity than with single π-pulses, as observed when comparing Fig.**11d** with Figs.**11b** and **c**. Furthermore, for both PRESTO with single π-pulses and RINEPT experiments, the maximal signal intensity is obtained for $\nu_1$ values higher than the theoretical ones: 161, 147 and 177 instead of 120, 135, and 160 kHz, for $SR4_1^2$, $R18_4^1$ and $R16_3^2$ recoupling, respectively. In fact, owing to the rf-field inhomogeneity in the sample space, higher signal can be obtained when the field at the center of the rotor exceeds the theoretical value.

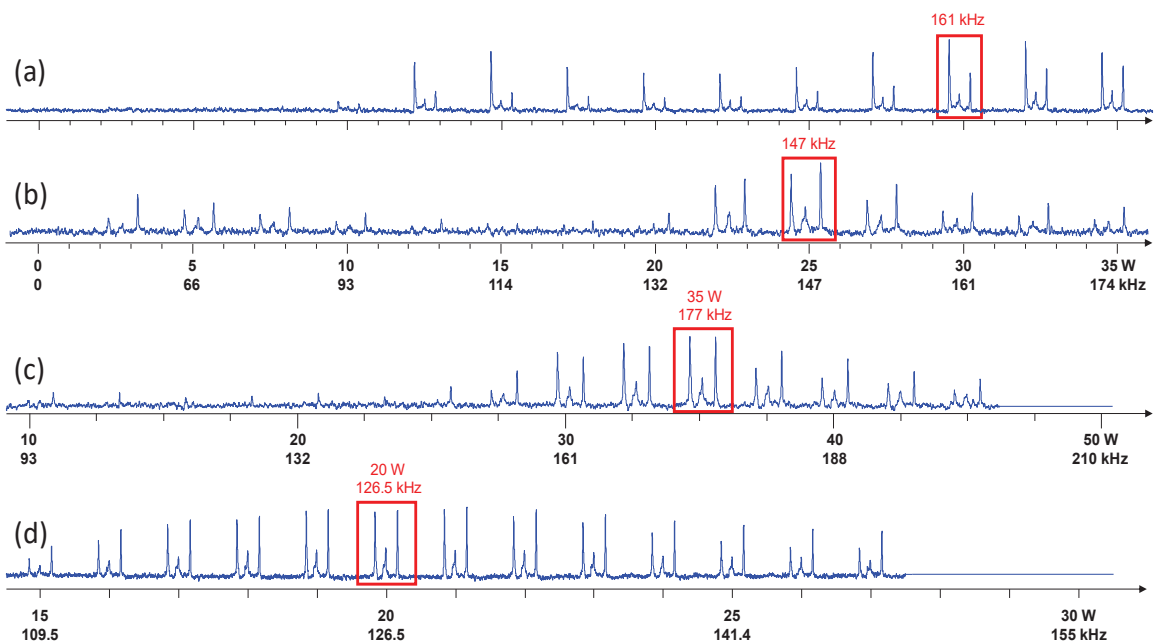

**Fig.11.** $^1$H → $^{27}$Al spectra of AlPO$_4$-14 versus $\nu_1$ at 18.8 T with RINEPT (a), and PRESTO with either single (b,c) or composite π-pulses (d), with $\nu_R$ = 60 kHz, $\tau$ = 800 µs and NS = 32. With PRESTO, the recoupling scheme is (b) $R18_4^1$, (c) $R16_3^2$ or (d) $R16_7^6$-C. It must be noted that the horizontal scales of (c) and (d) differ from that of (a) and (b).

Table 3 compares the S/N ratios of the cross-peaks in $^{27}$Al-{$^1$H} 2D spectra recorded at $B_0$ = 18.8 T with $\nu_R$ = 60 kHz using PRESTO-$R16_3^2$ with single π-pulses, PRESTO-$R16_7^6$-C and RINEPT. The PRESTO-$R16_3^2$ variant was chosen since it exhibits a slightly higher transfer efficiency than PRESTO-$R18_4^1$. Fig.**9c** displays the $^{27}$Al-{$^1$H} PRESTO-$R16_7^6$-C 2D spectrum of AlPO$_4$-14. As expected, the linewidth of the cross-peaks along the $^1$H dimension is about three times lower at $\nu_R$ = 60 than at 20 kHz, the $B_0$ field being constant (compare Figs.**9b** and **c**). Even if the acquisition times of the 2D spectra are 42 times longer at $\nu_R$ = 60 than at 20 kHz, their S/N are much smaller. Such sensitivity decrease stems from the smaller sample volume in 1.3 mm rotor with respect to 3.2 mm one. Table 3 indicates that RINEPT is in average 1.8 and 1.3-fold more sensitive than PRESTO-R16$_3^2$ and PRESTO-R16$_7^6$-C, respectively, at $B_0$ = 18.8 T with $\nu_R$ = 60 kHz. Under such condition, RINEPT exhibits the highest sensitivity since it is much more robust to rf-inhomogeneity, especially at high MAS frequencies (see Figs.**S5**, **S6** and **11**). Furthermore, we have shown that our 1.3 mm HX probe suffers from poor rf-field homogeneity [81]. Interestingly, PRESTO-R16$_7^6$-C method is more sensitive than PRESTO-R16$_3^2$ at 18.8 T with $\nu_R$ = 60 kHz, whereas it is less sensitive than PRESTO-R18$_2^5$ at 9.4 T with $\nu_R$ = 15 kHz. Such difference stems from (i) the higher robustness to rf-inhomogeneity of PRESTO-R16$_7^6$-C owing to the use of $270_0 90_{180}$ $\pi$-pulses, (ii) the decrease of non-averaged $^1$H-$^1$H interactions at ultra-fast MAS, and (iii) the increased robustness to these interactions of PRESTO-R16$_7^6$-C for higher MAS frequency, and hence higher rf-field of the pulses.

**Table 3.** S/N ratios at 18.8 T with AVANCE-IV console and $\nu_R$ = 60 kHz of the cross-peaks in $^{27}$Al-{$^1$H} 2D HETCOR spectra of AlPO$_4$-14 with RINEPT, PRESTO-R16$_3^2$ with single $\pi$-pulses, and PRESTO-R16$_7^6$-C.

| Sequence | PRESTO-R16$_3^2$ | | | RINEPT | | | PRESTO-R16$_7^6$-C | | |
|---|---|---|---|---|---|---|---|---|---|
| $\delta_{iso,27Al}$ / $\delta_{iso,1H}$ | 42 Al$^{IV}$ | 22 Al$^V$ | -2 Al$^{VI}$ | 42 Al$^{IV}$ | 22 Al$^V$ | -2 Al$^{VI}$ | 42 Al$^{IV}$ | 22 Al$^V$ | -2 Al$^{VI}$ |
| 1.3 (CH$_3$) | 56 | 10 | 21 | 121 | 14 | 39 | 111 | 18 | 29 |
| 3.3 (CH) | 12 | 2 | 3 | 28 | 3 | 7 | 21 | 3 | 4 |
| 5.8 (NH$_3^+$) | 37 | 30 | 75 | 85 | 36 | 111 | 63 | 34 | 70 |

## V.4. Experiments on $^{17}$O labeled fumed silica at 18.8 T with $\nu_R$ = 18 kHz

1D MAS spectra of $^{17}$O-labelled fumed silica were recorded at $B_0$ = 18.8 T with $\nu_R$ = 18 kHz and they are shown in Fig.**12**. The direct excitation spectrum is shown in Fig.**12a**. The de-shielded resonance is assigned to $^{17}$O nuclei in siloxane bridges, whereas the shielded one is assigned to $^{17}$O nuclei of silanol groups. As seen in Fig.**12b**, the use of QCPMG detection improves the sensitivity for the $^{17}$O siloxane signal. However, the silanol signal is then absent owing to its short $T_2$' constant time since the dipolar coupling between $^1$H and $^{17}$O nuclei leads to a rapid decay of the maximum of the echo signals during the QCPMG scheme. Conversely, the siloxane $^{17}$O nuclei, which are not bonded to protons, exhibit longer $T_2$' value, hence allowing the acquisition of 16 echoes.

No signal was detected with $^1$H $\rightarrow$ $^{17}$O CPMAS experiments for this sample. Such lack of signal illustrates the difficulty to optimize the CPMAS experiment when the S/N ratio is low, as it is the case for this sample. Conversely, as seen in Figs.**12c** and **d**, signals were detected for $^1$H $\rightarrow$ $^{17}$O RINEPT and PRESTO-R18$_2^5$ with single $\pi$-pulses. Both spectra exhibit signals for $^{17}$O siloxane and silanol nuclei; the last signal being more intense than that of siloxane, whereas it is the reverse for the direct excitation of $^{17}$O spectra (compare Figs.**12a** with **c**,**d**). Such variation in signal intensity stems from the more efficient $^1$H $\rightarrow$ $^{17}$O magnetization transfer for silanol than for siloxane since the $^1$H-$^{17}$O distance is shorter for the former group than for the latter one. In addition, the S/N ratio of the $^1$H $\rightarrow$ $^{17}$O RINEPT spectrum of fumed silica is 50 % higher than that of PRESTO-R18$_2^5$, whereas for AlPO$_4$-14, more efficient $^1$H $\rightarrow$ $^{27}$Al transfers were achieved at $\nu_R$ = 20 kHz using PRESTO, instead of RINEPT. RINEPT exhibits comparable robustness to rf-field inhomogeneity for both samples since (i) the strength of the $^1$H-$^1$H dipolar interactions determines the robustness to rf-field inhomogeneity of RINEPT (see Fig.**6b**) and (ii) the protons of fumed silica at $\nu_R$ = 18 kHz and AlPO$_4$-14 at $\nu_R$ = 20 kHz [not shown] exhibit NMR signals of comparable widths and hence, are subject to comparable $^1$H-$^1$H dipolar interactions. Furthermore, the shorter optimal $\tau$ delay for PRESTO in the case of fumed silica with respect to AlPO$_4$-14 (222 and 800 $\mu$s) indicates a larger $^1$H-S dipolar coupling for the former sample, which should increase the robustness to rf-inhomogeneity of PRESTO. Therefore, the inversion of the relative efficiencies of RINEPT and PRESTO in fumed silica with respect to AlPO$_4$-14 does not stem from a change in the robustness to rf inhomogeneity but from the higher robustness to CSA$_H$ of RINEPT with respect to PRESTO (Fig.**4**) since the $^1$H spectrum of the sample is dominated by a resonance at 2.9 ppm, typical of hydrogen-bonded silanols [18], which can be subject to CSA$_H$ as large as 30 ppm, i.e. 24 kHz at 18.8 T [77]. Conversely the $^1$H CSAs in AlPO$_4$-14 are expected to be smaller than 18 and 7 ppm for NH$_3^+$ and aliphatic protons, respectively [78].

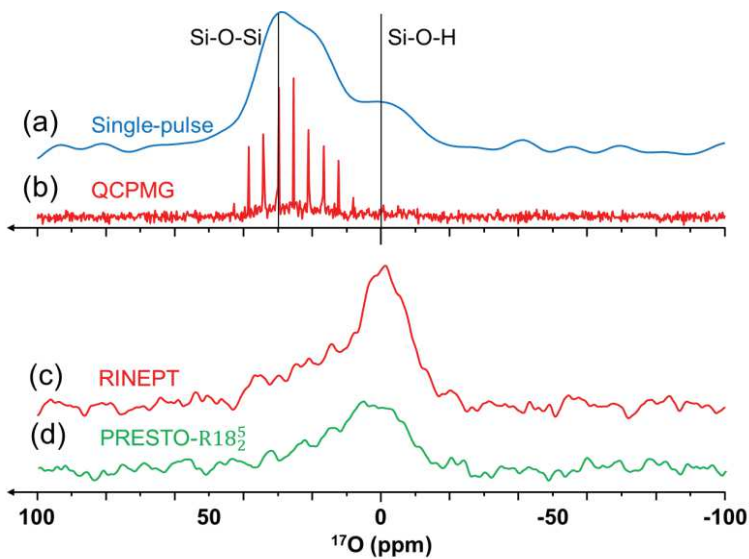

**Fig.12**. $^{17}$O MAS 1D spectra of labelled fumed silica at 18.8 T with $\nu_R$ = 18 kHz. Direct excitation recorded with (a) single-pulse and (b) QCPMG recycling with 16 echoes. $^1$H → $^{17}$O (c) RINEPT and (d) PRESTO-R18$_2^5$ with single π-pulses, τ = 222 μs and $\nu_1$ = 36 and 81 kHz for RINEPT and PRESTO-R18$_2^5$, respectively. The assignment of $^{17}$O resonances is displayed on the top. Spectra (a) and (b) were recorded with NS = 4,000 and τ$_{RD}$ = 0.5 s, i.e. $T_{exp}$ = 34 min, whereas the spectra (c) and (d) were recorded with *NS* = 40,000 and τ$_{RD}$ = 1 s, i.e. $T_{exp}$ = 11 h 7 min.

## VI. Conclusion

We have compared the performances of PRESTO and RINEPT experiments to transfer the magnetization of protons to half-integer spin quadrupolar isotopes. These two methods use different types of schemes: PRESTO employs γ-encoded $\mathrm{RN}_n^\nu$ recouplings, whereas RINEPT uses the $\mathrm{SR4}_1^2$ scheme, which is not. The reported simulations and experiments indicate that these techniques complement each other. Owing to its γ-encoding, PRESTO yields the highest transfer efficiency at low MAS frequency and low $B_0$ field, especially in the case of limited $^1$H offset and CSA. Conversely, RINEPT benefits from the highest transfer efficiency at high MAS frequency and high $B_0$ field, owing to its higher robustness to rf-field inhomogeneity, offset and CSA. In particular, RINEPT is beneficial for the transfer of polarization from protons subject to large CSA, such as those involved in hydrogen bonds, and notably will be useful for DNP experiments at high $B_0$ field, in which the DNP-enhanced magnetization of protons involved in hydrogen bonds has to be transferred to quadrupolar nuclei [23]. At high MAS frequency, the transfer efficiency of PRESTO can be enhanced by the use of composite 270$_0$90$_{180}$ π-pulses within the $\mathrm{R16}_7^6$-based recoupling scheme. These composite pulses improve the robustness of PRESTO to rf-field inhomogeneity. Nevertheless, the transfer efficiency of this PRESTO-R16$_7^6$-C technique remains smaller than that of RINEPT at high MAS frequency and high $B_0$ field. Furthermore, the robustness to $^1$H-$^1$H dipolar coupling of this PRESTO-R16$_7^6$-C sequence is not sufficient at low MAS frequencies. PRESTO sequences employing composite pulses suitable for low MAS frequencies, and hence DNP, are currently under investigation and will be presented elsewhere.

**Acknowledgements.** The authors would like to thank the reviewer who suggested the use of composite π-pulses in the PRESTO recoupling sequences. The Chevreul Institute (FR 2638), Ministère de l'Enseignement Supérieur et de la Recherche, Région Hauts-de-France and FEDER are acknowledged for supporting and funding partially this work. Authors also acknowledge contract CEFIPRA n°85208-E, PRC CNRS-NSFC, ANR-14-CE07-0009-01 and ANR-17-ERC2-0022 (EOS). This project has also received funding from the European Union's Horizon 2020 research and innovation program under grant agreement n°731019 (EUSMI). Authors would like to thank Drs. Régis Gauvin and Tom Vamcompernolle for synthesizing $^{17}$O-labelled fumed silica.

[1] S.E. Ashbrook, S. Sneddon, New Methods and Applications in Solid-State NMR Spectroscopy of Quadrupolar Nuclei, J. Am. Chem. Soc. 136 (2014) 15440–15456. doi:10.1021/ja504734p.

[2] M.T. Janicke, C.C. Landry, S.C. Christiansen, D. Kumar, G.D. Stucky, B.F. Chmelka, Aluminum Incorporation and Interfacial Structures in MCM-41 Mesoporous Molecular Sieves, J. Am. Chem. Soc. 120 (1998) 6940–6951. doi:10.1021/ja972633s.

[3] M.D. Alba, M.A. Castro, M. Naranjo, A.C. Perdigón, Structural localization of $Al^{3+}$ ions in aluminosilicates: application of heteronuclear chemical shift correlation to 2:1 phyllosilicates, Phys. Chem. Miner. 31 (2004) 195–202. doi:10.1007/s00269-003-0361-z.

[4] X. Xue, M. Kanzaki, Al coordination and water speciation in hydrous aluminosilicate glasses: Direct evidence from high-resolution heteronuclear $^1$H–$^{27}$Al correlation NMR, Solid State Nucl. Magn. Reson. 31 (2007) 10–27. doi:10.1016/j.ssnmr.2006.11.001.

[5] B. Bouchevreau, C. Martineau, C. Mellot-Draznieks, A. Tuel, M.R. Suchomel, J. Trébosc, O. Lafon, J.-P. Amoureux, F. Taulelle, High-Resolution Structural Characterization of Two Layered Aluminophosphates by Synchrotron Powder Diffraction and NMR Crystallographies, Chem Mater. 25 (2013) 2227–2242.

[6] L. Mafra, J. Rocha, C. Fernandez, F.A. Almeida Paz, Characterization of microporous aluminophosphate IST-1 using $^1$H Lee-Goldburg techniques., J. Magn. Reson. 180 (2006) 236–44. doi:10.1016/j.jmr.2006.02.017.

[7] J. Wack, R. Siegel, T. Ahnfeldt, N. Stock, L. Mafra, J. Senker, Identifying Selective Host–Guest Interactions Based on Hydrogen Bond Donor–Acceptor Pattern in Functionalized Al-MIL-53 Metal–Organic Frameworks, J. Phys. Chem. C. 117 (2013) 19991–20001. doi:10.1021/jp4063252.

[8] G. Tricot, J. Trébosc, F. Pourpoint, R.M. Gauvin, L. Delevoye, The D-HMQC MAS-NMR Technique: An Efficient Tool for the Editing of Through-Space Correlation Spectra Between Quadrupolar and Spin-1/2, Annu. Rep. NMR Spectrosc. 81 (2014) 145–184. doi:10.1016/B978-0-12-800185-1.00004-8.

[9] M. Taoufik, K.C. Szeto, N. Merle, I.D. Rosal, L. Maron, J. Trébosc, G. Tricot, R.M. Gauvin, L. Delevoye, Heteronuclear NMR Spectroscopy as a Surface-Selective Technique: A Unique Look at the Hydroxyl Groups of γ-Alumina., Chem. – Eur. J. 20 (2014) 4038–4046. doi:10.1002/chem.201304883.

[10] A. Gallo, A. Fong, K.C. Szeto, J. Rieb, L. Delevoye, R.M. Gauvin, M. Taoufik, B. Peters, S.L. Scott, Ligand Exchange-Mediated Activation and Stabilization of a Re-Based Olefin

Metathesis Catalyst by Chlorinated Alumina, J. Am. Chem. Soc. 138 (2016) 12935–12947. doi:10.1021/jacs.6b06953.

[11] M.A. Bashir, T. Vancompernolle, R.M. Gauvin, L. Delevoye, N. Merle, V. Monteil, M. Taoufik, T.F.L. McKenna, C. Boisson, Silica/MAO/(n-BuCp)$_2$ZrCl$_2$ catalyst: effect of support dehydroxylation temperature on the grafting of MAO and ethylene polymerization, Catal. Sci. Technol. 6 (2016) 2962–2974. doi:10.1039/C5CY01285F.

[12] Z. Wang, Y. Jiang, O. Lafon, J. Trébosc, K.D. Kim, C. Stampfl, A. Baiker, J.-P. Amoureux, J. Huang, Brønsted acid sites based on penta-coordinated aluminum species, Nat. Commun. 7 (2016) 13820. doi:10.1038/ncomms13820.

[13] E. Dib, T. Mineva, E. Veron, V. Sarou-Kanian, F. Fayon, B. Alonso, ZSM-5 Zeolite: Complete Al Bond Connectivity and Implications on Structure Formation from Solid-State NMR and Quantum Chemistry Calculations, J. Phys. Chem. Lett. 9 (2018) 19–24. doi:10.1021/acs.jpclett.7b03050.

[14] K.C. Szeto, A. Gallo, S. Hernández-Morejudo, U. Olsbye, A. De Mallmann, F. Lefebvre, R.M. Gauvin, L. Delevoye, S.L. Scott, M. Taoufik, Selective Grafting of Ga(i-Bu)$_3$ on the Silanols of Mesoporous H-ZSM-5 by Surface Organometallic Chemistry, J. Phys. Chem. C. 119 (2015) 26611–26619. doi:10.1021/acs.jpcc.5b09289.

[15] S.-J. Hwang, C.-Y. Chen, S.I. Zones, Boron Sites in Borosilicate Zeolites at Various Stages of Hydration Studied by Solid State NMR Spectroscopy, J. Phys. Chem. B. 108 (2004) 18535–18546. doi:10.1021/jp0476904.

[16] A. Wong, D. Laurencin, R. Dupree, M.E. Smith, Two-dimensional $^{43}$Ca–$^1$H correlation solid-state NMR spectroscopy, Solid State Nucl. Magn. Reson. 35 (2009) 32–36. doi:10.1016/j.ssnmr.2008.11.002.

[17] D. Lee, C. Leroy, C. Crevant, L. Bonhomme-Coury, F. Babonneau, D. Laurencin, C. Bonhomme, G.D. Paëpe, Interfacial Ca$^{2+}$ environments in nanocrystalline apatites revealed by dynamic nuclear polarization enhanced $^{43}$Ca NMR spectroscopy, Nat. Commun. 8 (2017) 14104. doi:10.1038/ncomms14104.

[18] N. Merle, J. Trébosc, A. Baudouin, I.D. Rosal, L. Maron, K. Szeto, M. Genelot, A. Mortreux, M. Taoufik, L. Delevoye, R.M. Gauvin, $^{17}$O NMR Gives Unprecedented Insights into the Structure of Supported Catalysts and Their Interaction with the Silica Carrier, J. Am. Chem. Soc. 134 (2012) 9263–9275. doi:10.1021/ja301085m.

[19] F.A. Perras, U. Chaudhary, I.I. Slowing, M. Pruski, Probing Surface Hydrogen Bonding and Dynamics by Natural Abundance, Multidimensional, $^{17}$O DNP-NMR Spectroscopy, J. Phys. Chem. C. 120 (2016) 11535–11544. doi:10.1021/acs.jpcc.6b02579.

[20] F.G. Vogt, H. Yin, R.G. Forcino, L. Wu, $^{17}$O Solid-State NMR as a Sensitive Probe of Hydrogen Bonding in Crystalline and Amorphous Solid Forms of Diflunisal, Mol. Pharm. 10 (2013) 3433–3446. doi:10.1021/mp400275w.

[21] E.G. Keeler, V.K. Michaelis, M.T. Colvin, I. Hung, P.L. Gor'kov, T.A. Cross, Z. Gan, R.G.

Griffin, $^{17}$O MAS NMR Correlation Spectroscopy at High Magnetic Fields, J. Am. Chem. Soc. 139 (2017) 17953–17963. doi:10.1021/jacs.7b08989.

[22] F. Blanc, L. Sperrin, D.A. Jefferson, S. Pawsey, M. Rosay, C.P. Grey, Dynamic Nuclear Polarization Enhanced Natural Abundance $^{17}$O Spectroscopy, J. Am. Chem. Soc. 135 (2013) 2975–2978. doi:10.1021/ja4004377.

[23] F.A. Perras, T. Kobayashi, M. Pruski, Natural Abundance $^{17}$O DNP Two-Dimensional and Surface-Enhanced NMR Spectroscopy, J. Am. Chem. Soc. 137 (2015) 8336–8339. doi:10.1021/jacs.5b03905.

[24] D.A. Hirsh, A.J. Rossini, L. Emsley, R.W. Schurko, $^{35}$Cl dynamic nuclear polarization solid-state NMR of active pharmaceutical ingredients, Phys. Chem. Chem. Phys. 18 (2016) 25893–25904. doi:10.1039/C6CP04353D.

[25] M.K. Pandey, H. Kato, Y. Ishii, Y. Nishiyama, Two-dimensional proton-detected $^{35}$Cl/$^{1}$H correlation solid-state NMR experiment under fast magic angle sample spinning: application to pharmaceutical compounds, Phys. Chem. Chem. Phys. 18 (2016) 6209–6216. doi:10.1039/C5CP06042G.

[26] Z. Gan, J.-P. Amoureux, J. Trébosc, Proton-detected $^{14}$N MAS NMR using homonuclear decoupled rotary resonance, Chem. Phys. Lett. 435 (2007) 163–169. doi:10.1016/j.cplett.2006.12.066.

[27] S. Cavadini, A. Abraham, G. Bodenhausen, Proton-detected nitrogen-14 NMR by recoupling of heteronuclear dipolar interactions using symmetry-based sequences, Chem. Phys. Lett. 445 (2007) 1–5. doi:10.1016/j.cplett.2007.07.060.

[28] A.L. Webber, S. Masiero, S. Pieraccini, J.C. Burley, A.S. Tatton, D. Iuga, T.N. Pham, G.P. Spada, S.P. Brown, Identifying guanosine self assembly at natural isotopic abundance by high-resolution $^{1}$H and $^{13}$C solid-state NMR spectroscopy., J. Am. Chem. Soc. 133 (2011) 19777–19795. doi:10.1021/ja206516u.

[29] G.N.M. Reddy, D.S. Cook, D. Iuga, R.I. Walton, A. Marsh, S.P. Brown, An NMR crystallography study of the hemihydrate of 2', 3'-O-isopropylideneguanosine, Solid State Nucl. Magn. Reson. 65 (2015) 41–48. doi:10.1016/j.ssnmr.2015.01.001.

[30] S.P. Brown, Nitrogen–Proton Correlation Experiments of Organic Solids at Natural Isotopic Abundance, in: eMagRes, John Wiley & Sons, Ltd, 2014. doi:10.1002/9780470034590.emrstm1323.

[31] A.S. Tatton, T.N. Pham, F.G. Vogt, D. Iuga, A.J. Edwards, S.P. Brown, Probing intermolecular interactions and nitrogen protonation in pharmaceuticals by novel $^{15}$N-edited and 2D $^{14}$N-$^{1}$H solid-state NMR, CrystEngComm. 14 (2012) 2654–2659. doi:10.1039/c2ce06547a.

[32] K. Maruyoshi, D. Iuga, O.N. Antzutkin, S.P. Velaga, S.P. Brown, Identifying the intermolecular hydrogen-bonding supramolecular synthons in an indomethacin – nicotinamide cocrystal by solid-state NMR, Chem. Commun. 48 (2012) 10844–10846.

doi:10.1039/c2cc36094b.

[33]   A.S. Tatton, T.N. Pham, F.G. Vogt, D. Iuga, A.J. Edwards, S.P. Brown, Probing Hydrogen Bonding in Cocrystals and Amorphous Dispersions Using $^{14}$N–$^{1}$H HMQC Solid-State NMR, Mol. Pharm. 10 (2013) 999–1007.

[34]   T. Venâncio, L.M. Oliveira, J. Ellena, N. Boechat, S.P. Brown, Probing intermolecular interactions in a diethylcarbamazine citrate salt by fast MAS $^{1}$H solid-state NMR spectroscopy and GIPAW calculations, Solid State Nucl. Magn. Reson. 87 (2017) 73–79. doi:10.1016/j.ssnmr.2017.02.006.

[35]   S.L. Veinberg, K.E. Johnston, M.J. Jaroszewicz, B.M. Kispal, C.R. Mireault, T. Kobayashi, M. Pruski, R.W. Schurko, Natural abundance $^{14}$N and $^{15}$N solid-state NMR of pharmaceuticals and their polymorphs, Phys. Chem. Chem. Phys. 18 (2016) 17713–17730. doi:10.1039/C6CP02855A.

[36]   O. Lafon, Q. Wang, B. Hu, F. Vasconcelos, J. Trébosc, S. Cristol, F. Deng, J.-P. Amoureux, Indirect Detection via Spin-1/2 Nuclei in Solid State NMR Spectroscopy: Application to the Observation of Proximities between Protons and Quadrupolar Nuclei, J. Phys. Chem. A. 113 (2009) 12864–12878. doi:10.1021/jp906099k.

[37]   C.A. Fyfe, H. Grondey, K.T. Mueller, K.C. Wong-Moon, T. Markus, Coherence transfer involving quadrupolar nuclei in solids: aluminum-27 - phosphorus-31 cross-polarization NMR in the molecular sieve VPI-5, J. Am. Chem. Soc. 114 (1992) 5876–5878. doi:10.1021/ja00040a069.

[38]   A.J. Vega, CP/MAS of quadrupolar S = 3/2 nuclei., Solid State Nucl. Magn. Reson. 1 (1992) 17–32.

[39]   J.-P. Amoureux, M. Pruski, Theoretical and experimental assessment of single-and multiple-quantum cross-polarization in solid state NMR, Mol. Phys. 100 (2002) 1595–1613. doi:10.1080/0026897021012575.

[40]   A.J. Vega, MAS NMR spin locking of half-integer quadrupolar nuclei, J. Magn. Reson. 1969. 96 (1992) 50–68. doi:10.1016/0022-2364(92)90287-H.

[41]   S.E. Ashbrook, S. Wimperis, Spin-locking of half-integer quadrupolar nuclei in nuclear magnetic resonance of solids: second-order quadrupolar and resonance offset effects., J. Chem. Phys. 131 (2009) 194509. doi:10.1063/1.3263904.

[42]   G. Tricot, O. Lafon, J. Trébosc, L. Delevoye, F. Méar, L. Montagne, J.-P. Amoureux, Structural characterisation of phosphate materials: new insights into the spatial proximities between phosphorus and quadrupolar nuclei using the D-HMQC MAS NMR technique, Phys. Chem. Chem. Phys. 13 (2011) 16786–16794. doi:10.1039/C1CP20993K.

[43]   C.A. Fyfe, K.T. Mueller, H. Grondey, K.C. Wong-Moon, Dipolar dephasing between quadrupolar and spin- nuclei. REDOR and TEDOR NMR experiments on VPI-5, Chem. Phys. Lett. 199 (1992) 198–204. doi:10.1016/0009-2614(92)80069-N.

[44]   J. Trébosc, B. Hu, J.-P. Amoureux, Z. Gan, Through-space $R^3$-HETCOR experiments

between spin-1/2 and half-integer quadrupolar nuclei in solid-state NMR, J. Magn. Reson. 186 (2007) 220–227. doi:10.1016/j.jmr.2007.02.015.

[45]   C. Martineau, B. Bouchevreau, F. Taulelle, J. Trébosc, O. Lafon, J.-P. Amoureux, High-resolution through-space correlations between spin-1/2 and half-integer quadrupolar nuclei using the MQ-D-R-INEPT NMR experiment., Phys. Chem. Chem. Phys. 14 (2012) 7112–7119. doi:10.1039/c2cp40344g.

[46]   A. Venkatesh, M.P. Hanrahan, A.J. Rossini, Proton detection of MAS solid-state NMR spectra of half-integer quadrupolar nuclei, Solid State Nucl. Magn. Reson. 84 (2017) 171–181. doi:10.1016/j.ssnmr.2017.03.005.

[47]   X. Zhao, W. Hoffbauer, J. Schmedt auf der Günne, M.H. Levitt, Heteronuclear polarization transfer by symmetry-based recoupling sequences in solid-state NMR., Solid State Nucl. Magn. Reson. 26 (2004) 57–64. doi:10.1016/j.ssnmr.2003.11.001.

[48]   F.A. Perras, T. Kobayashi, M. Pruski, PRESTO polarization transfer to quadrupolar nuclei: implications for dynamic nuclear polarization, Phys. Chem. Chem. Phys. 17 (2015) 22616–22622. doi:10.1039/C5CP04145G.

[49]   J.D. van Beek, R. Dupree, M.H. Levitt, Symmetry-based recoupling of $^{17}$O-$^{1}$H spin pairs in magic-angle spinning NMR., J. Magn. Reson. 179 (2006) 38–48. doi:10.1016/j.jmr.2005.11.003.

[50]   A.H. Hing, S. Vega, J. Schaefer, Transferred-Echo Double-Resonance NMR, 209 (1992) 205–209.

[51]   T. Gullion, J. Schaefer, Rotational-echo double-resonance NMR, J Magn Reson. 81 (1989) 196–200. doi:10.1016/j.jmr.2011.09.003.

[52]   A. Brinkmann, A.P.M. Kentgens, Proton-Selective $^{17}$O-$^{1}$H Distance Measurements in Fast Magic-Angle-Spinning Solid-State NMR Spectroscopy for the Determination of Hydrogen Bond Lengths, J Am Chem Soc. 128 (2006) 14758–14759.

[53]   A. Venkatesh, M.J. Ryan, A. Biswas, K.C. Boteju, A.D. Sadow, A.J. Rossini, Enhancing the Sensitivity of Solid-State NMR Experiments with Very Low Gyromagnetic Ratio Nuclei with Fast Magic Angle Spinning and Proton Detection, J. Phys. Chem. A. 122 (2018) 5635–5643. doi:10.1021/acs.jpca.8b05107.

[54]   M.K. Pandey, M. Malon, A. Ramamoorthy, Y. Nishiyama, Composite-180° pulse-based symmetry sequences to recouple proton chemical shift anisotropy tensors under ultrafast MAS solid-state NMR spectroscopy, J. Magn. Reson. 250 (2015) 45–54. doi:10.1016/j.jmr.2014.11.002.

[55]   Z. Gan, $^{13}$C/$^{14}$N heteronuclear multiple-quantum correlation with rotary resonance and REDOR dipolar recoupling., J. Magn. Reson. 184 (2007) 39–43. doi:10.1016/j.jmr.2006.09.016.

[56]   Z. Gan, J.-P. Amoureux, J. Trébosc, Proton-detected $^{14}$N MAS NMR using homonuclear decoupled rotary resonance, Chem. Phys. Lett. 435 (2007) 163–169.

doi:10.1016/j.cplett.2006.12.066.

[57]  B. Hu, J. Trébosc, J.-P. Amoureux, Comparison of several hetero-nuclear dipolar recoupling NMR methods to be used in MAS HMQC/HSQC., J. Magn. Reson. 192 (2008) 112–122. doi:10.1016/j.jmr.2008.02.004.

[58]  X. Lu, O. Lafon, J. Trébosc, G. Tricot, L. Delevoye, F. Méar, L. Montagne, J.-P. Amoureux, Observation of proximities between spin-1/2 and quadrupolar nuclei: which heteronuclear dipolar recoupling method is preferable?, J. Chem. Phys. 137 (2012) 144201. doi:10.1063/1.4753987.

[59]  H. Nagashima, A.S. Lilly Thankamony, J. Trébosc, L. Montagne, G. Kerven, J.-P. Amoureux, O. Lafon, Observation of proximities between spin-1/2 and quadrupolar nuclei in solids: Improved robustness to chemical shielding using adiabatic symmetry-based recoupling, Solid State Nucl. Magn. Reson. 94 (2018) 7–19. doi:10.1016/j.ssnmr.2018.07.001.

[60]  H. Nagashima, A.S. Lilly Thankamony, J. Trébosc, F. Pourpoint, O. Lafon, J.-P. Amoureux, γ-Independent through-space hetero-nuclear correlation between spin-1/2 and quadrupolar nuclei in solids, Solid State Nucl. Magn. Reson. 84 (2017) 216–226. doi:10.1016/j.ssnmr.2017.06.002.

[61]  Y. Ishii, R. Tycko, Sensitivity Enhancement in Solid State [15]N NMR by Indirect Detection with High-Speed Magic Angle Spinning, J. Magn. Reson. 142 (2000) 199–204. doi:10.1006/jmre.1999.1976.

[62]  G. Pileio, M. Concistrè, N. McLean, A. Gansmüller, R.C.D. Brown, M.H. Levitt, Analytical theory of γ-encoded double-quantum recoupling sequences in solid-state nuclear magnetic resonance, J. Magn. Reson. 186 (2007) 65–74. doi:10.1016/j.jmr.2007.01.009.

[63]  L. Mafra, S.M. Santos, R. Siegel, I. Alves, F.A. Almeida Paz, D. Dudenko, H.W. Spiess, Packing Interactions in Hydrated and Anhydrous Forms of the Antibiotic Ciprofloxacin: a Solid-State NMR, X-ray Diffraction, and Computer Simulation Study, J. Am. Chem. Soc. 134 (2012) 71–74. doi:10.1021/ja208647n.

[64]  M. Sardo, R. Siegel, S.M. Santos, J. Rocha, J.R.B. Gomes, L. Mafra, Combining Multinuclear High-Resolution Solid-State MAS NMR and Computational Methods for Resonance Assignment of Glutathione Tripeptide, J. Phys. Chem. A. 116 (2012) 6711–6719. doi:10.1021/jp302128r.

[65]  D. Marion, M. Ikura, R. Tschudin, A. Bax, Rapid recording of 2D NMR spectra without phase cycling. Application to the study of hydrogen exchange in proteins, J. Magn. Reson. 1969. 85 (1989) 393–399. doi:10.1016/0022-2364(89)90152-2.

[66]  M. Bak, J.T. Rasmussen, N.C. Nielsen, SIMPSON: a general simulation program for solid-state NMR spectroscopy, J. Magn. Reson. 147 (2000) 296–330.

[67]  M. Bak, N.C. Nielsen, REPULSION, a novel approach to efficient powder averaging in solid-state NMR, J. Magn. Reson. 125 (1997) 132.

[68]  LAPACK — Linear Algebra PACKage, (n.d.). http://www.netlib.org/lapack/ (accessed

October 28, 2018).

[69]   Z. Tošner, R. Andersen, B. Stevensson, M. Edén, N.C. Nielsen, T. Vosegaard, Computer-intensive simulation of solid-state NMR experiments using SIMPSON, J. Magn. Reson. 246 (2014) 79–93. doi:10.1016/j.jmr.2014.07.002.

[70]   M. Veshtort, R.G. Griffin, SPINEVOLUTION: a powerful tool for the simulation of solid and liquid state NMR experiments, J. Magn. Reson. 178 (2006) 248–282. doi:10.1016/j.jmr.2005.07.018.

[71]   M. Deschamps, Chapter Three - Ultrafast Magic Angle Spinning Nuclear Magnetic Resonance, in: G.A. Webb (Ed.), Annu. Rep. NMR Spectrosc., Academic Press, 2014: 109–144. doi:10.1016/B978-0-12-800185-1.00003-6.

[72]   C.A. Fyfe, H. Meyer zu Altenschildesche, K.C. Wong-Moon, H. Grondey, J.-M. Chezeau, 1D and 2D solid state NMR investigations of the framework structure of As-synthesized $AlPO_4$-14., Solid State Nucl. Magn. Reson. 9 (1997) 97–106.

[73]   S. Björgvinsdóttir, B.J. Walder, A.C. Pinon, J.R. Yarava, L. Emsley, DNP enhanced NMR with flip-back recovery, J. Magn. Reson. 288 (2018) 69–75. doi:10.1016/j.jmr.2018.01.017.

[74]   S.R. Chaudhari, D. Wisser, A.C. Pinon, P. Berruyer, D. Gajan, P. Tordo, O. Ouari, C. Reiter, F. Engelke, C. Copéret, M. Lelli, A. Lesage, L. Emsley, Dynamic Nuclear Polarization Efficiency Increased by Very Fast Magic Angle Spinning, J. Am. Chem. Soc. 139 (2017) 10609–10612. doi:10.1021/jacs.7b05194.

[75]   F.H. Larsen, H.J. Jakobsen, P.D. Ellis, N.C. Nielsen, Sensitivity-Enhanced Quadrupolar-Echo NMR of Half-Integer Quadrupolar Nuclei. Magnitudes and Relative Orientation of Chemical Shielding and Quadrupolar Coupling Tensors, J. Phys. Chem. A. 101 (1997) 8597–8606.

[76]   D.H. Brouwer, J.A. Ripmeester, Symmetry-based recoupling of proton chemical shift anisotropies in ultrahigh-field solid-state NMR., J. Magn. Reson. 185 (2007) 173–8. doi:10.1016/j.jmr.2006.12.003.

[77]   L. Duma, D. Abergel, P. Tekely, G. Bodenhausen, Proton chemical shift anisotropy measurements of hydrogen-bonded functional groups by fast magic-angle spinning solid-state NMR spectroscopy., Chem. Commun. 3 (2008) 2361–3. doi:10.1039/b801154k.

[78]   L. Liang, G. Hou, X. Bao, Measurement of proton chemical shift anisotropy in solid-state NMR spectroscopy, Solid State Nucl. Magn. Reson. 93 (2018) 16–28. doi:10.1016/j.ssnmr.2018.04.002.

[79]   R. Gupta, G. Hou, T. Polenova, A.J. Vega, RF inhomogeneity and how it controls CPMAS, Solid State Nucl. Magn. Reson. 72 (2015) 17–26. doi:10.1016/j.ssnmr.2015.09.005.

[80]   R.N. Purusottam, G. Bodenhausen, P. Tekely, Effects of inherent rf field inhomogeneity on heteronuclear decoupling in solid-state NMR, Chem. Phys. Lett. 635 (2015) 157–162. doi:10.1016/j.cplett.2015.06.051.

[81]    H. Nagashima, J. Trébosc, O. Lafon, F. Pourpoint, P. Paluch, M.J. Potrzebowski, J.-P. Amoureux, Imaging the spatial distribution of radiofrequency field, sample and temperature in MAS NMR rotor, Solid State Nucl. Magn. Reson. 87 (2017) 137–142. doi:10.1016/j.ssnmr.2017.08.001.

[82]    S. Odedra, S. Wimperis, Imaging of the $B_1$ distribution and background signal in a MAS NMR probehead using inhomogeneous $B_0$ and $B_1$ fields, J. Magn. Reson. 231 (2013) 95–99. doi:10.1016/j.jmr.2013.04.002.

[83]    Z. Tošner, A. Purea, J.O. Struppe, S. Wegner, F. Engelke, S.J. Glaser, B. Reif, Radiofrequency fields in MAS solid state NMR probes, J. Magn. Reson. 284 (2017) 20–32. doi:10.1016/j.jmr.2017.09.002.

[84]    M. Carravetta, M. Edén, X. Zhao, A. Brinkmann, M.H. Levitt, Symmetry principles for the design of radiofrequency pulse sequences in the nuclear magnetic resonance of rotating solids, Chem Phys Lett. 321 (2000) 205–215.

[85]    M.H. Levitt, Encyclopedia of Nuclear Magnetic Resonance, in: Wiley, Chichester, 2002: 165–196.

[86]    L. Delevoye, C. Fernandez, C.M. Morais, J.-P. Amoureux, V. Montouillout, J. Rocha, Double-resonance decoupling for resolution enhancement of $^{31}$P solid-state MAS and $^{27}$Al → $^{31}$P MQHETCOR NMR., Solid State Nucl. Magn. Reson. 22 (2002) 501–512. doi:10.1006/snmr.2002.0080.

[87]    H. Nagashima, J. Trébosc, L. Calvez, F. Pourpoint, F. Mear, O. Lafon, J.-P. Amoureux, $^{71}$Ga-$^{77}$Se connectivities and proximities in gallium selenide crystal and glass probed by solid-state NMR, J. Magn. Reson. 282 (2017) 71–82. doi:10.1016/j.jmr.2017.07.009.

# Supplementary Information

# Magnetization transfer from protons to quadrupolar nuclei in solid-state NMR using PRESTO or dipolar-mediated refocused INEPT methods

Raynald Giovine,[1] Julien Trébosc,[1] Frédérique Pourpoint,[1] Olivier Lafon,[1,2*] Jean-Paul Amoureux[1,3*]

[1] Univ. Lille, CNRS-8181, UCCS: Unit of Catalysis and Chemistry of Solids, F-59000 Lille, France.

[2] IUF, Institut Universitaire de France, 1 rue Descartes, 75231 Paris, France.

[3] Bruker France, 34 rue de l'Industrie, F-67166 Wissembourg, France.

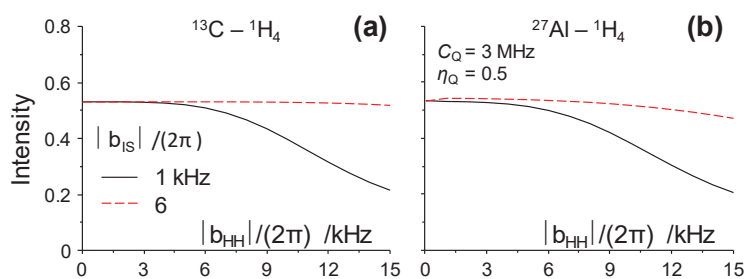

**Fig.S1.** Simulated on-resonance RINEPT transfer efficiency at 18.8 T and $\nu_R$ = 60 kHz for fifteen $b_{HH}$ values, with $CSA_H$ = 0 and $|b_{IS}|/(2\pi)$ = 1 or 6 kHz for (a) $^{13}C$-$^1H_4$ and (b) $^{27}Al$-$^1H_4$ spin-systems. The total computing time was of 30 h for (b), whereas it was only of 1 h for (a). The $\tau$ value was set to its optimum value determined from Fig.S2a in (a) and from similar build-up curves simulated for $^{27}Al$-$^1H$ spin-pair in (b) [not shown].

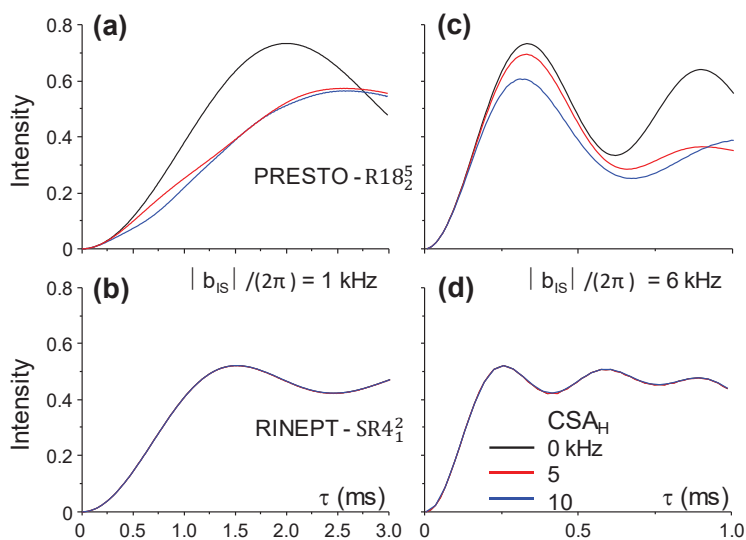

**Fig.S2.** Simulated build-up curves of $^1H \rightarrow ^{13}C$ transfer at 18.8 T and $\nu_R$ = 60 kHz of (a,c) PRESTO-R$18_2^5$ with single $\pi$-pulses or (b,d) RINEPT for an isolated $^{13}C$-$^1H$ spin pair with $|b_{IS}|/(2\pi)$ = (a,b) 1 or (c,d) 6 kHz, and $CSA_H$ = 0, 5 or 10 kHz.

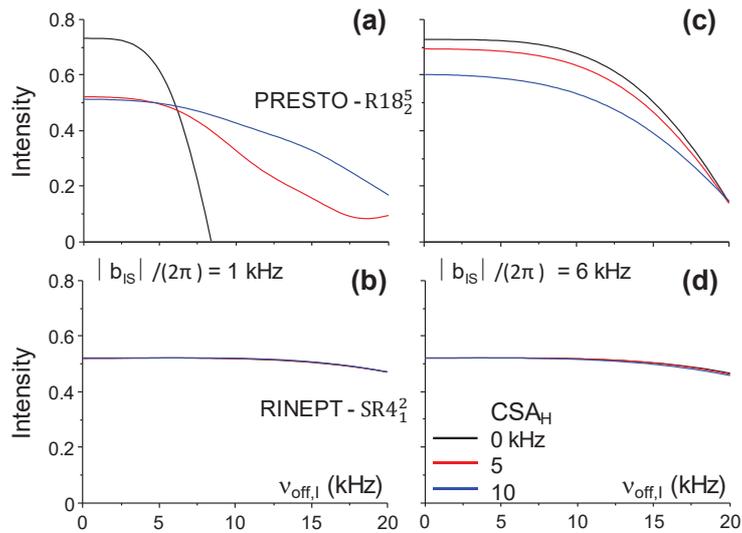

**Fig.S3.** Simulated transfer efficiency versus the $^1$H resonance offset, $\nu_{off,I}$, at 18.8 T and $\nu_R$ = 60 kHz for (a,c) PRESTO-R$18_2^5$ with single π-pulses or (b,d) RINEPT, with $|b_{IS}|/(2\pi)$ = (a,b) 1 or (c,d) 6 kHz, and CSA$_H$ = 0, 5 or 10 kHz. The τ value was set to its optimum value determined from Fig.S2.

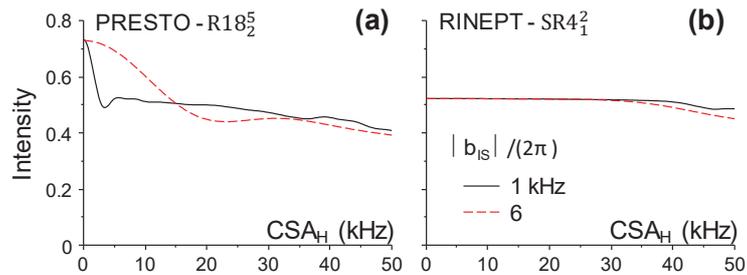

**Fig.S4.** Simulated transfer efficiency versus CSA$_H$ at 18.8 T and $\nu_R$ = 60 kHz for (a) PRESTO-R$18_2^5$ with single π-pulses and (b) RINEPT with $|b_{IS}|/(2\pi)$ = 1 or 6 kHz. The τ value was set to its optimum value determined from Fig.S2.

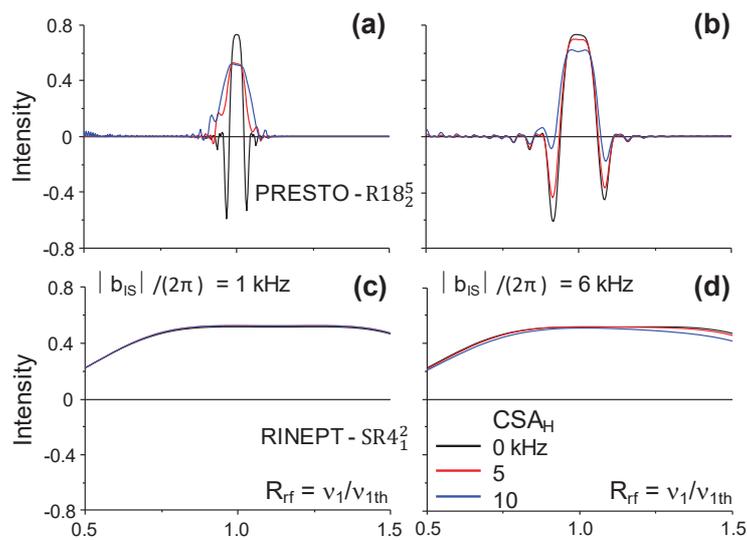

**Fig.S5.** Simulated transfer efficiency at 18.8 T and $\nu_R$ = 60 kHz versus the rf-inhomogeneity, $R_{rf} = \nu_1/\nu_{1th}$, for (a,b) PRESTO-R$18_2^5$ with single π-pulses or (c,d) RINEPT, with CSA$_I$ = 0, 5, 10 and $|b_{IS}|/(2\pi)$ = 1 (a,c) or 6 (b,d) kHz. The τ value was set to its optimum value determined from Fig.S2.

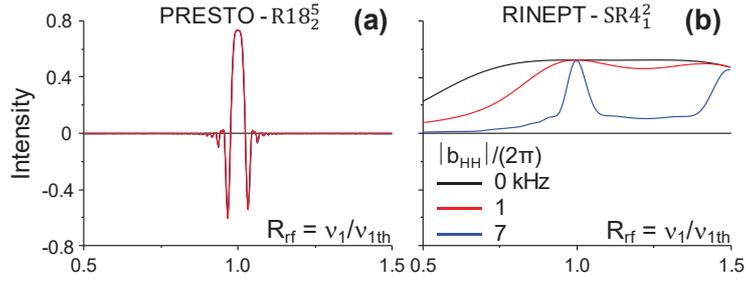

**Fig.S6.** Simulated transfer efficiency at 18.8 T and $\nu_R$ = 60 kHz versus the rf-inhomogeneity, $R_{rf} = \nu_1/\nu_{1th}$, for (a) PRESTO-$R18_2^5$ with single π-pulses or (b) RINEPT sequences applied to $^{13}$C-$^1$H$_4$ spin system with $|b_{HH}|/(2\pi)$ = 0, 1 or 7 kHz, $|b_{IS}|/(2\pi)$ = 1 kHz and CSA$_H$ = 0. The τ value was set to its optimum value determined from Fig.S2. Note the different vertical scalings.

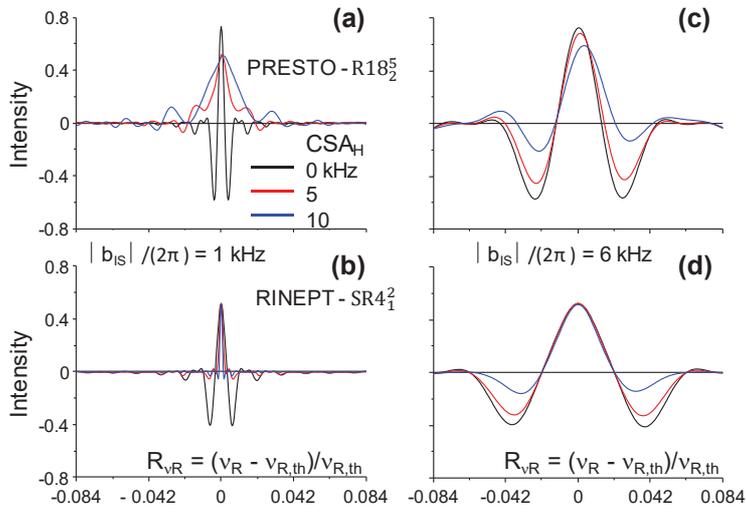

**Fig.S7.** Simulated transfer efficiency at 18.8 T and $\nu_{R,th} \approx 23$ kHz versus $R_{\nu R} = (\nu_R - \nu_{R,th})/\nu_{R,th}$ for (a,c) PRESTO-$R18_2^5$ with single π-pulses or (b,d) RINEPT transfers with $|b_{IS}|/(2\pi)$ = (a,b) 1 or (c,d) 6 kHz and CSA$_H$ = 0, 5 or 10 kHz. The τ value was set to its optimum value determined from Fig.2.

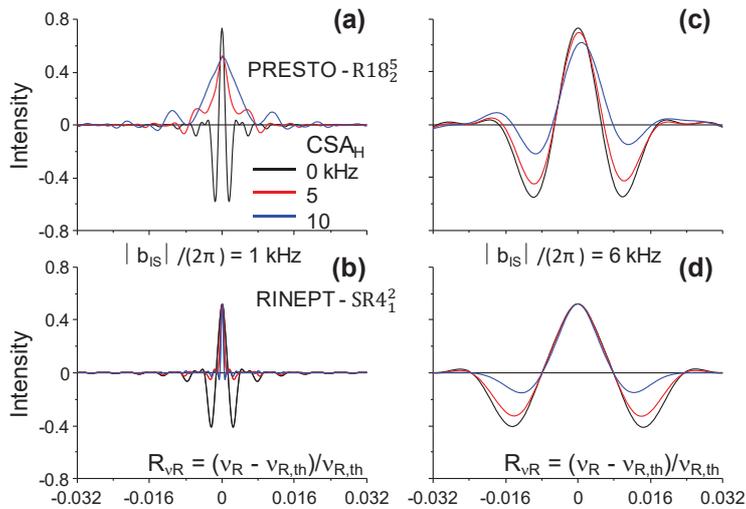

**Fig.S8.** Simulated transfer efficiency at 18.8 T and $\nu_{R,th}$ = 60 kHz versus $R_{\nu R} = (\nu_R - \nu_{R,th})/\nu_{R,th}$ for (a,c) PRESTO-$R18_2^5$ with single π-pulses or (b,d) RINEPT techniques with $|b_{IS}|/(2\pi)$ = (a,b) 1 or (c,d) 6 kHz and CSA$_H$ = 0, 5 or 10 kHz. The τ value was set to its optimum value determined from Fig.S2. Note the horizontal scales are smaller than in Fig.S7 (0.032 % 0.084).

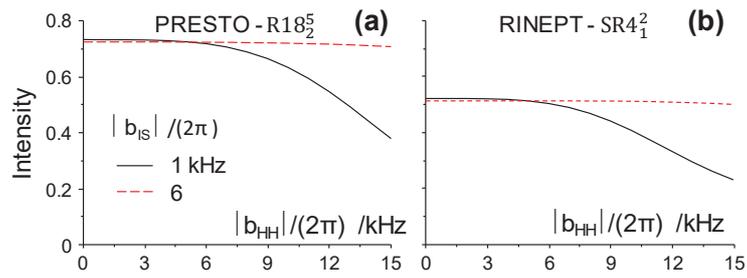

**Fig.S9.** Simulated on-resonance transfer efficiency at 18.8 T and $\nu_R$ = 60 kHz versus the $|b_{HH}|/(2\pi)$ constant in $^{13}$C-$^{1}$H$_4$ spin system for (a) PRESTO-R$18_2^5$ with single π-pulses and (b) RINEPT, with CSA$_H$ = 0 and $|b_{IS}|/(2\pi)$ = 1 or 6 kHz. The τ value was set to its optimum value determined from Fig.S2.

**Full phase cycle of the various sequences.**

PRESTO-III-R$18_2^5$:

| $(R18_2^5)_0$ | $(R18_2^5)_{180}$ | $(R18_2^5)_{90}$ | $(R18_2^5)_{270}$ |
|---|---|---|---|
| 50° | 230° | 140° | 320° |
| 310° | 130° | 40° | 220° |

*D*-RINEPT-SR$4_1^2$:

| SR$4_1^2$ | | |
|---|---|---|
| Phase Shift 0° | Phase Shift 120° | Phase Shift 240° |
| 90° | 210° | 330° |
| 270° | 30° | 150° |
| 90° | 210° | 330° |
| 270° | 30° | 150° |
| 270° | 30° | 150° |
| 90° | 210° | 330° |
| 270° | 30° | 150° |
| 90° | 210° | 330° |

**Pulse programs:**

## PRESTO-III-RN$_n^\nu$

```
;PRESTO-III.jt
; for topspin 3
; tested AlPO-14 : 2D OK
; PRESTO-III polarization transfer
; version 1.0 (published online XXX)
; ----------------
; DESCRIPTION :
; PRESTO-III experiment using R1817 or R1825
; AUTHOR
; Hiroki Nagashima / Julien TREBOSC
;reference sequence: PRESTO-II.hn
;article: Zhao X Solid State Nucl Magn Reson. 2004 Sep;26(2):57-64.
;Heteronuclear polarization transfer by symmetry-based recoupling sequences in solid-state NMR

;$COMMENT=PRESTO-III polarization transfer
;$CLASS=Solids
;$DIM=1D
;$TYPE=
;$SUBTYPE=
;$OWNER=Hiroki

; ------------
;PARAMETERS:
;ns... see below in phase cycling section
;d1 : recycle delay
;d4: Z filter delay
;pl21 : RF power level p4 and p5
;pl23 : recoupling power level
;pl2 : power level for p2
;p2 : 90 degree pulse @ pl2
;p4 : 90 degree pulse @ pl21
;p5 : 180 degree pulse @ pl21
;p6 : recoupling 180 degree pulse  @ pl23
;p16 : T1 dipolar recoupling time [in us]
;p26 : T2 dipolar recoupling time [in us]
;p17 : actual dip. rec. time
;cnst31 : =MAS spin rate
;no p1 : 90 degree pulse @pl3
;p3 : 90 degree pulse @ pl10
;pl10 : p3 power
;pl2 : p2 power

;FnMODE : States or States-TPPI
;
;ZGOPTNS : PRESATf1 PRESATf3 decF3 decF2t1 decF2aq Distance _R1817 _R1825
; PRESAT : send presaturation pulses on F1 can be replaced with DS=1 or 2
; decF3 : applyies decoupling during aq on F3
; decF2aq : applyies decoupling during aq on F2 (1H)
; decF2t1 : applyies decoupling during t1 on F2 (1H)

;********************* PRESAT ****************
#include "presat.incl"
#ifndef PRESATf2
#undef PRESAT2
#define PRESAT2(f2)
#endif

#ifndef PRESATf1
#undef PRESAT1
#define PRESAT1(f1)
#endif

#ifdef decF3
#define dec
#define decF3on cpds3:f3
#define decF3off do:f3
#else
#define decF3on
#define decF3off
#endif

#ifdef decF2aq
#define decF2
#define decF2aqon cpds3:f2
#else
#define decF2aqon
#endif

#ifdef decF2
#include "decouple.incl"
#define tppm
#define decF2off do:f2
#else
#define decF2off
#endif

;************* calculation of t1 delays *****************

#define ZGOPNTS_ERROR

#ifndef _R1817
#ifndef _R1825
#undef ZGOPTNS_ERROR
#define ZGOPTNS_ERROR you must use ZGOPTNS -D_R1817 or -D_R1825
#endif
#endif
ZGOPNTS_ERROR

#ifdef _R1817
;this is R1817 symmetry
"p6=(1.0/18)*1s/cnst31"
"l23=trunc(((p16/2)/p6)+0.5)"      ; +0.5 will round to nearest integer
#ifdef Distance
"l24=trunc(((p26/2)/p6)+0.5)"      ; +0.5 will round to nearest integer
#endif
#define RN_p_phi_p_0 7000
#define RN_m_phi_p_0 29000
#define RN_p_phi_p_90 16000
#define RN_m_phi_p_90 2000
#define RN_p_phi_p_180 25000
#define RN_m_phi_p_180 11000
#define RN_p_phi_p_270 34000
#define RN_m_phi_p_270 20000
#endif

#ifdef _R1825
;this is R1825 symmetry
"p6=(2.0/18)*1s/cnst31"
"l23=trunc(((p16/2)/p6)+0.5)"      ; +0.5 will round to nearest integer
#ifdef Distance
"l24=trunc(((p26/2)/p6)+0.5)"      ; +0.5 will round to nearest integer
#endif
#define RN_p_phi_p_0 5000
```

```
#define RN_m_phi_p_0 31000
#define RN_p_phi_p_90 14000
#define RN_m_phi_p_90 4000
#define RN_p_phi_p_180 23000
#define RN_m_phi_p_180 13000
#define RN_p_phi_p_270 32000
#define RN_m_phi_p_270 22000
#endif

define delay nTr
define delay delA
define delay delB
define delay Tr
define delay delC
define delay delD
define delay Dmin
define loopcounter Lmin

#ifdef _R1431
;this is R1431 symmetry
"p6=(3.0/14)*1s/cnst31"
"l23=trunc((p16/p6)/2 +0.5)"   ; +0.5 will round to nearest integer
#ifdef Distance
"l24=trunc((p26/p6)/2 +0.5)"   ; +0.5 will round to nearest integer
#endif
#define RN_p_phi_p_0 1286
#define RN_m_phi_p_0 3471
#define RN_p_phi_p_90 10286
#define RN_m_phi_p_90 7714
#define RN_p_phi_p_180 19286
#define RN_m_phi_p_180 16714
#define RN_p_phi_p_270 28286
#define RN_m_phi_p_270 26714
#endif

#ifdef _R1841
;this is R1841 symmetry
"p6=(4.0/18)*1s/cnst31"
"l23=trunc((p16/p6)/2 +0.5)"   ; +0.5 will round to nearest integer
#ifdef Distance
"l24=trunc((p26/p6)/2 +0.5)"   ; +0.5 will round to nearest integer
#endif
#define RN_p_phi_p_0 1000
#define RN_m_phi_p_0 35000
#define RN_p_phi_p_90 10000
#define RN_m_phi_p_90 8000
#define RN_p_phi_p_180 19000
#define RN_m_phi_p_180 17000
#define RN_p_phi_p_270 28000
#define RN_m_phi_p_270 27000
#endif

#ifdef _R1632
;this is R1841 symmetry
"p6=(3.0/16)*1s/cnst31"
"l23=trunc((p16/p6)/2 +0.5)"   ; +0.5 will round to nearest integer
#ifdef Distance
"l24=trunc((p26/p6)/2 +0.5)"   ; +0.5 will round to nearest integer
#endif
#define RN_p_phi_p_0 2250
#define RN_m_phi_p_0 33750
#define RN_p_phi_p_90 11250
#define RN_m_phi_p_90 6750
#define RN_p_phi_p_180 20250
#define RN_m_phi_p_180 15750
#define RN_p_phi_p_270 29250
#define RN_m_phi_p_270 24750
#endif
"Tr=1/cnst31"
"p5=2*p4"

"p17=2*l23*p6"
#ifdef Distance
"p27=2*l24*p6"
#endif

"delA=p6*l23-p5/2"
"delB=p6*l23-p4/2-p5/2"

"in0=inf1"

;************** experiment block *******************

1 ze
"p17=2*l23*p6"
2 100m decF2off decF3off
rpp21
rpp22
rpp23
rpp24
ip21*18000
ip22*18000
  PRESAT2(f2)
  d1
  PRESAT1(f1)

(p2 pl2 ph3):f2
d0
(p2 pl2 ph4):f2
d4 pl23:f2; Z filter
exec_on_chan:t1:f2
RN_1, (p6 ph21 ipp21 ipp22):f2 ;PRESTO T1 recoupling phase 0
  lo to RN_1 times l23
RN_12, (p6 ph22 ipp21 ipp22):f2 ;PRESTO T1 recoupling phase 0+180
  lo to RN_12 times l23

RN_2, (p6 ph23 ipp23 ipp24):f2 ;PRESTO T2 recoupling phase 90
#ifdef Distance
  lo to RN_2 times l24
#else
  lo to RN_2 times l23
#endif
RN_22, (p6 ph24 ipp23 ipp24):f2 ;PRESTO T2 recoupling phase 90+180
#ifdef Distance
  lo to RN_22 times l24
#else
  lo to RN_22 times l23
#endif
exec_wait

exec_on_other
delA pl21:f1
(p5 ph0 ):f1
delB
(p4 ph1 ):f1
delB
(p5 ph2):f1
exec_wait

  go=2 ph31 decF2aqon decF3on ;1H decoupling on
  10u decF3off decF2off
  100m mc #0 to 2 F1PH(ip3, id0)
; for 1D version
;  100m mc #0 to 2 F0(zd)
exit

;phase cycling n*8
```

```
ph1={0 0}
ph2=0
ph0=0
ph3= {{0 0}}^2
ph4=2
ph21=(36000) {RN_p_phi_p_0  RN_m_phi_p_0}
ph22=(36000) {RN_p_phi_p_180 RN_m_phi_p_180}
ph23=(36000) {RN_p_phi_p_90  RN_m_phi_p_90}
ph24=(36000) {RN_p_phi_p_270 RN_m_phi_p_270}
ph31={{0 2}}^2
;set phases for presat : ph19 and ph20
presatPH
```

## PRESTO-III-RN$_n^v$-C

```
;PRESTO-III.jt
; for topspin 3

; PRESTO-III polarization transfer
; version 1.0 (published online XXX)
; ----------------
; DESCRIPTION :
; PRESTO-III experiment using R1817 or R1825 or R1841
;AUTHOR
; Hiroki Nagashima / Julien TREBOSC
;reference sequence: PRESTO-II.hn
;article: Zhao X Solid State Nucl Magn Reson. 2004
Sep;26(2):57-64.
;Heteronuclear polarization transfer by symmetry-based
recoupling sequences in solid-state NMR

;$COMMENT=PRESTO-III polarization transfer
;$CLASS=Solids
;$DIM=1D
;$TYPE=
;$SUBTYPE=
;$OWNER=Hiroki

; -----------
;PARAMETERS:
;ns... see below in phase cycling section
;d1 : recycle delay
;d4: zFilter delay (20us)
;pl21 : RF power level p4 and p5
;pl23 : recoupling power level
;p2 : 90 degree pulse @ pl2
;p4 : 90 degree pulse @ pl21
;p5 : 180 degree pulse @ pl21
;p6 : recoupling 360 degree pulse  @ pl23
;p16 : T1 dip. rec. time [us] (up to p90 on F1)
;p26 : T2 dip. rec. time [us] (from p90 on F1)
;p17 : actual rec. time after rounding p16
;cnst31 : =MAS spin rate
;no p1 : 90 degree pulse @pl3
;p3 : 90 degree pulse @ pl10
;pl10 : p3 power
;pl2 : p2 power

;FnMODE : States or States-TPPI
;
;ZGOPTNS : PRESATf1 PRESATf3 decF3 decF2t1 decF2aq
Distance _R1817 _R1825
; PRESAT : send presaturation pulses on F1 can be
replaced with DS=1 or 2
; decF3 : applies decoupling during aq on F3
; decF2aq : applies decoupling during aq on F2 (1H)
; decF2t1 : applies decoupling during t1 on F2 (1H)

;*****************PRESAT********************
#include "presat.incl"
#ifndef PRESATf2
#undef PRESAT2
#define PRESAT2(f2)
#endif

#ifndef PRESATf1
#undef PRESAT1
#define PRESAT1(f1)
#endif

#ifdef decF3
#define dec
#define decF3on cpds3:f3
#define decF3off do:f3
#else
#define decF3on
#define decF3off
#endif

#ifdef decF2aq
#define decF2
#define decF2aqon cpds3:f2
#else
#define decF2aqon
#endif

#ifdef decF2
#include "decouple.incl"
#define tppm
#define decF2off do:f2
#else
#define decF2off
#endif

;***********calculation of t1 delays **************

#define ZGOPNTS_ERROR
#ifndef _R1817cp
#ifndef _R1825cp
```

```
#undef ZGOPTNS_ERROR
#define ZGOPTNS_ERROR you must use ZGOPTNS -D_R1817cp or -D_R1825cp
#endif
#endif
ZGOPNTS_ERROR

#ifdef _R1817cp
;this is R1817 symmetry with composite pulse
"p6=(1.0/18)*1s/cnst31"
"l23=trunc((p16/p6)/2 +0.5)"    ; +0.5 will round to nearest integer
#ifdef Distance
"l24=trunc((p26/p6)/2 +0.5)"    ; +0.5 will round to nearest integer
#endif
#define RN_p_phi_p_0 7000
#define RN_m_phi_p_0 29000
#define RN_p_phi_p_90 16000
#define RN_m_phi_p_90 2000
#define RN_p_phi_p_180 25000
#define RN_m_phi_p_180 11000
#define RN_p_phi_p_270 34000
#define RN_m_phi_p_270 20000
#endif

#ifdef _R1825cp
;this is R1825 symmetry with composite pulse
"p6=(2.0/18)*1s/cnst31"
"l23=trunc((p16/p6)/2+0.5)"   ; +0.5 will round to nearest integer
#ifdef Distance
"l24=trunc((p16/p6)/2+0.5)"   ; +0.5 will round to nearest integer
#endif
#define RN_p_phi_p_0 5000
#define RN_m_phi_p_0 31000
#define RN_p_phi_p_90 14000
#define RN_m_phi_p_90 4000
#define RN_p_phi_p_180 23000
#define RN_m_phi_p_180 13000
#define RN_p_phi_p_270 32000
#define RN_m_phi_p_270 22000
#endif

#ifdef _R1431cp
;this is R1431 symmetry with composite pulse
"p6=(3.0/14)*1s/cnst31"
"l23=trunc((p16/p6)/2 +0.5)"    ; +0.5 will round to nearest integer
#ifdef Distance
"l24=trunc((p26/p6)/2 +0.5)"    ; +0.5 will round to nearest integer
#endif
#define RN_p_phi_p_0 1286
#define RN_m_phi_p_0 3471
#define RN_p_phi_p_90 10286
#define RN_m_phi_p_90 7714
#define RN_p_phi_p_180 19286
#define RN_m_phi_p_180 16714
#define RN_p_phi_p_270 28286
#define RN_m_phi_p_270 26714
#endif

#ifdef _R1841cp
;this is R1841 symmetry with composite pulse
"p6=(4.0/18)*1s/cnst31"
"l23=trunc((p16/p6)/2 +0.5)"    ; +0.5 will round to nearest integer
#ifdef Distance
"l24=trunc((p26/p6)/2 +0.5)"    ; +0.5 will round to nearest integer
#endif
#define RN_p_phi_p_0 1000
#define RN_m_phi_p_0 35000
#define RN_p_phi_p_90 10000
#define RN_m_phi_p_90 8000
#define RN_p_phi_p_180 19000
#define RN_m_phi_p_180 17000
#define RN_p_phi_p_270 28000
#define RN_m_phi_p_270 27000
#endif

#ifdef _R1632cp
;this is R16_3^2 symmetry with composite pulse
"p6=(3.0/16)*1s/cnst31"
"l23=trunc((p16/p6)/2 +0.5)"    ; +0.5 will round to nearest integer
#ifdef Distance
"l24=trunc((p26/p6)/2 +0.5)"    ; +0.5 will round to nearest integer
#endif
#define RN_p_phi_p_0 2250
#define RN_m_phi_p_0 33750
#define RN_p_phi_p_90 11250
#define RN_m_phi_p_90 6750
#define RN_p_phi_p_180 20250
#define RN_m_phi_p_180 15750
#define RN_p_phi_p_270 29250
#define RN_m_phi_p_270 24750
#endif

#ifdef _R1696cp
;this is R16_9^6 symmetry with composite pulse
"p6=(9.0/16)*1s/cnst31"
"l23=trunc((p16/p6)/2 +0.5)"    ; +0.5 will round to nearest integer
#ifdef Distance
"l24=trunc((p26/p6)/2 +0.5)"    ; +0.5 will round to nearest integer
#endif
#define RN_p_phi_p_0 6750
#define RN_m_phi_p_0 29250
#define RN_p_phi_p_90 15750
#define RN_m_phi_p_90 2250
#define RN_p_phi_p_180 24750
#define RN_m_phi_p_180 11250
#define RN_p_phi_p_270 33750
#define RN_m_phi_p_270 20250
#endif

#ifdef _R1676cp
;this is R16_7^6 symmetry with composite pulse
"p6=(7.0/16)*1s/cnst31"
"l23=trunc((p16/p6)/2 +0.5)"    ; +0.5 will round to nearest integer
```

```
#ifdef Distance
"l24=trunc((p26/p6)/2 +0.5)"    ; +0.5 will round to nearest integer
#endif
#define RN_p_phi_p_0 6750
#define RN_m_phi_p_0 29250
#define RN_p_phi_p_90 15750
#define RN_m_phi_p_90 2250
#define RN_p_phi_p_180 24750
#define RN_m_phi_p_180 11250
#define RN_p_phi_p_270 33750
#define RN_m_phi_p_270 20250
#endif

define delay nTr
define delay delA
define delay delB
define delay Tr
define delay delC
define delay delD
define delay Dmin
define loopcounter Lmin

"Tr=1/cnst31"
"p5=2*p4"

"p17=2*l23*p6"
#ifdef Distance
"p27=2*l24*p6"
#endif

"delA=p6*l23-p5/2"
"delB=p6*l23-p4/2-p5/2"

"in0=inf1"
;*************experiment block ****************

1 ze
"p17=2*l23*p6"
2 100m decF2off decF3off
rpp21
rpp22
rpp23
rpp24
ip21*18000
ip22*18000
  PRESAT2(f2)
   d1
  PRESAT1(f1)

(p2 pl2 ph3):f2
d0
(p2 pl2 ph4):f2
d4 pl23:f2; Z filter
exec_on_chan:t1:f2
RN_1, (p6*0.75 ph21 ):f2 ;PRESTO T1 recoupling phase 0
    (p6*0.25 ph22 ipp21 ipp22):f2 ;PRESTO T1 recoupling phase 0
  lo to RN_1 times l23
RN_12, (p6*0.75 ph22 ):f2 ;PRESTO T1 recoupling phase 0+180
    (p6*0.25 ph21 ipp21 ipp22 ):f2 ;PRESTO T1 recoupling phase 0+180
  lo to RN_12 times l23

RN_2, (p6*0.75 ph23 ):f2 ;PRESTO T2 recoupling phase 90
    (p6*0.25 ph24 ipp23 ipp24):f2 ;PRESTO T2 recoupling phase 90
#ifdef Distance
 lo to RN_2 times l24
#else
 lo to RN_2 times l23
#endif
RN_22, (p6*0.75 ph24 ):f2 ;PRESTO T2 recoupling phase 90+180
    (p6*0.25 ph23 ipp23 ipp24):f2 ;PRESTO T2 recoupling phase 90+180
#ifdef Distance
 lo to RN_22 times l24
#else
 lo to RN_22 times l23
#endif
exec_wait

exec_on_other
delA pl21:f1
(p5 ph0 ):f1
delB
(p4 ph1 ):f1
delB
(p5 ph2):f1
exec_wait

 go=2 ph31 decF2aqon decF3on ;1H decoupling on
 10u decF3off decF2off
 100m mc #0 to 2 F1PH(ip3, id0)
; for 1D version
;  100m mc #0 to 2 F0(zd)
exit

;phase cycling n*8
ph1={0 0}
ph2=0
ph0=0
ph3= {{0 0}}^2
ph4=2
ph21=(36000) { RN_p_phi_p_0  RN_m_phi_p_0 }
ph22=(36000) { RN_p_phi_p_180 RN_m_phi_p_180 }
ph23=(36000) { RN_p_phi_p_90 RN_m_phi_p_90 }
ph24=(36000) { RN_p_phi_p_270 RN_m_phi_p_270 }
ph31={0 2}^2
;set phases for presat : ph19 and ph20
presatPH
```

*D*-RINEPT-SR4$_1^2$

```
;D-INEPT-SR4.jtmp
;avance-version (02/05/31)
;INEPT for non-selective polarization transfer
;with decoupling during acquisition
; MODIFICATIONS
; 08/05/2012 : JT added the d4 right after d0 for proper
echo formation

;d0 initial t1 evolution time (=0)
;d6 probe dead time (should be D6=DE)
;pl1 : p1 and p2 power level
;pl21 p3 and p4 power level
;pl11 dipolar recoupling power (sr4/sfam)
;spnam5 dipolar recoupling shape pulse
;sp5 power for recoupling shape
;l11 sr4/sfam repetition
;cnst30 sfam offset
;cnst31: spinning speed in Hz
;cnst3 sfam shape pulse step (ns)(400ns)
;pl12: power level for CPD/BB decoupling
;p1 90 degree high power pulse
;p2 180 degree high power pulse
;p3 90 degree high power pulse
;p4 180 degree high power pulse
;p6 pulse of the recoupling sequence
;d1 : relaxation delay; 1-5 * T1
;NS: 4 * n, total number of scans: NS * TD0
;DS: 16
;cpd1: decoupling during R3
;cpdprg1: decoupling during R3
;cpd2: decoupling during AQ and t1
;cpdprg2: decoupling during AQ and t1
;cpd3: decoupling during AQ
;cpdprg3: decoupling during AQ

#include <Avance.incl>

; storeVC option to store VClist used when popting MAS
#ifdef storeVC
#define VCstored vclab, 1u \n lo to vclab times c
#else
#define VCstored
#endif

;-)))))))
#include "presat.incl"
;-)
#ifndef PRESATf2
#undef PRESAT2
#define PRESAT2(f2)
#endif
;-)
#ifndef PRESATf1
#undef PRESAT1
#define PRESAT1(f1)
#endif
;-(((((((
;---------- DECOUPLING ---------------
#include "decouple.incl"

#ifdef decF2
#define decF2off do:f2
#define decF2aqon cpds2:f2
#else
#define decF2aqon
#define decF2off
#endif

;..............SFAM/SR4......................
#ifndef _SR4
#define _SFAM
#endif

define delay RF
define delay dummy

#ifdef _SR4
"p6=0.25s/cnst31"
"l11=trunc((p16/p6)/8+0.5)"   ; +0.5 will round to nearest integer
"p17=2*p6*4*l11"
"RF=500000/p6"
"dummy=RF+p17"
#endif

#ifdef _SFAM
"p6=1s/cnst31"
"l11=trunc((p16/p6)/2+0.5)"   ; +0.5 will round to nearest integer
"p17=2*p6*l11"
"dummy=p17+cnst30+cnst3"
#endif
;...........................................

"p2=p1*2"

"p4=p3*2"
"d6=de"
"d2=0.5s/(cnst31)-p3/2.0-d0-0.5u"
"d3=0.5s/(cnst31)-p4/2.0-0.5u"
"d4=0.5s/(cnst31)-p3/2.0-0.5u"
"d5=0.5s/(cnst31)-d6"
"in0=inf1"

define delay showInAsed
"showInAsed=cnst3+dummy"

1 ze
VCstored
; protection
if "p2+2*d3+4u>p4" goto pass
print "p4 too large"
goto HaltAQ
pass, 1m
"showInAsed=cnst3+dummy"

2 30m decF2off
  PRESAT2(f2)
  d1 rpp16 rpp17 rpp14 rpp15 ; not necessary to use different phases and reset but...
  PRESAT1(f1)
```

```
  (10u pl21):f2 (10u pl1 ph2):f1
  (p3 ph1):f2
  d0
  d2 ;fq=cnst23:f1
  0.5u pl11:f2

#ifdef _SFAM
;can be R3 if no modulation at all
SFAMl1, (p6:sp5 (currentpower) ph6):f2
lo to SFAMl1 times l11
#endif
#ifdef _SR4
sr4_1, (p6 ph16^):f2
  (p6 ph16^):f2
  (p6 ph16^):f2
  (p6 ph16^):f2

  lo to sr4_1 times l11
#endif

  (center (d3 0.5u pl21 ph2 p4 ph2 d3 0.5u pl11 ph6):f2
(p2 ph0):f1 )

#ifdef _SFAM
;can be R3 if not modulation at all
SFAMl2, (p6:sp5 (currentpower) ph7):f2
lo to SFAMl2 times l11
#endif
#ifdef _SR4
sr4_2, (p6 ph17^):f2
  (p6 ph17^):f2
  (p6 ph17^):f2
  (p6 ph17^):f2

  lo to sr4_2 times l11
#endif

(center (d4 0.5u pl21 ph3 p3 ph3 d4 0.5u pl11 ph6):f2
(p1 ph0):f1 )

#ifdef _SFAM
;can be R3 if not modulation at all
SFAMl3, (p6:sp5 (currentpower) ph6):f2
lo to SFAMl3 times l11
#endif
#ifdef _SR4
sr4_3, (p6 ph15^):f2
  (p6 ph15^):f2
  (p6 ph15^):f2
  (p6 ph15^):f2

  lo to sr4_3 times l11
#endif

(center (d3 0.5u pl21 ph4 p4 ph4 d3 0.5u pl11 ph6):f2
(p2 ph0):f1 )

#ifdef _SFAM
;can be R3 if not modulation at all
SFAMl4, (p6:sp5 (currentpower) ph7):f2
lo to SFAMl4 times l11
#endif
#ifdef _SR4
sr4_4, (p6 ph14^):f2
  (p6 ph14^):f2
  (p6 ph14^):f2
  (p6 ph14^):f2

  lo to sr4_4 times l11
#endif

  d5 ;fq=0:f1
  go=2 ph31 decF2aqon
  10u decF2off
  30m mc #0 to 2 F1PH(ip1,id0)

HaltAQ, 1m

exit

ph0=0
ph2=0
ph4=1
ph6=0
ph7=0
ph3=1
ph16= (360) 90 270 90 270 270 90 270 90 210 30 210 30
30 210 30 210 330 150 330 150 150 330 150 330
ph17= (360) 90 270 90 270 270 90 270 90 210 30 210 30
30 210 30 210 330 150 330 150 150 330 150 330
ph15= (360) 90 270 90 270 270 90 270 90 210 30 210 30
30 210 30 210 330 150 330 150 150 330 150 330
ph14= (360) 90 270 90 270 270 90 270 90 210 30 210 30
30 210 30 210 330 150 330 150 150 330 150 330

ph10=0

#ifdef opt1D
ph1=0 2 1 3
ph31=0 2 1 3
#else
ph1=0 2
ph31=0 2
#endif
presatPH
```